# Networks of Influence Diagrams: A Formalism for Representing Agents' Beliefs and Decision-Making Processes

**Ya'akov Gal**                                        GAL@CSAIL.MIT.EDU
*MIT Computer Science and Artificial Intelligence Laboratory*
*Harvard School of Engineering and Applied Sciences*

**Avi Pfeffer**                                        AVI@EECS.HARVARD.EDU
*Harvard School of Engineering and Applied Sciences*

## Abstract

This paper presents Networks of Influence Diagrams (NID), a compact, natural and highly expressive language for reasoning about agents' beliefs and decision-making processes. NIDs are graphical structures in which agents' mental models are represented as nodes in a network; a mental model for an agent may itself use descriptions of the mental models of other agents. NIDs are demonstrated by examples, showing how they can be used to describe conflicting and cyclic belief structures, and certain forms of bounded rationality. In an opponent modeling domain, NIDs were able to outperform other computational agents whose strategies were not known in advance. NIDs are equivalent in representation to Bayesian games but they are more compact and structured than this formalism. In particular, the equilibrium definition for NIDs makes an explicit distinction between agents' optimal strategies, and how they actually behave in reality.

## 1. Introduction

In recent years, decision theory and game theory have had a profound impact on the design of intelligent systems. Decision theory provides a mathematical language for single-agent decision-making under uncertainty, whereas game theory extends this language to the multi-agent case. On a fundamental level, both approaches provide a definition of what it means to build an intelligent agent, by equating intelligence with utility maximization. Meanwhile, graphical languages such as Bayesian networks (Pearl, 1988) have received much attention in AI because they allow for a compact and natural representation of uncertainty in many domains that exhibit structure. These formalisms often lead to significant savings in representation and in inference time (Dechter, 1999; Cowell, Lauritzen, & Spiegelhater, 2005).

Recently, a wide variety of representations and algorithms have augmented graphical languages to be able to represent and reason about agents' decision-making processes. For the single-agent case, influence diagrams (Howard & Matheson, 1984) are able to represent and to solve an agent's decision making problem using the principles of decision theory. This representation has been extended to the multi-agent case, in which decision problems are solved within a game-theoretic framework (Koller & Milch, 2001; Kearns, Littman, & Singh, 2001).

The focus in AI so far has been on the classical, normative approach to decision and game theory. In the classical approach, a game specifies the actions that are available to the agents, as well as their utilities that are associated with each possible set of agents' actions. The game is then analyzed to determine rational strategies for each of the agents. Fundamental to this approach are the assumptions that the structure of the game, including agents' utilities and their actions, is known to all of the agents, that agents' beliefs about the game are consistent with each other and correct, that all agents reason about the game in the same way, and that all agents are rational in that they choose the strategy that maximizes their expected utility given their beliefs.

As systems involving multiple, autonomous agents become ubiquitous, they are increasingly deployed in open environments comprising human decision makers and computer agents that are designed by or represent different individuals or organizations. Examples of such systems include on-line auctions, and patient care-delivery systems (MacKie-Mason, Osepayshivili, Reeves, & Wellman, 2004; Arunachalam & Sadeh, 2005). These settings are challenging because no assumptions can be made about the decision-making strategies of participants in open environments. Agents may be uncertain about the structure of the game or about the beliefs of other agents about the structure of the game; they may use heuristics to make decisions or they may deviate from their optimal strategies (Camerer, 2003; Gal & Pfeffer, 2003b; Rajarshi, Hanson, Kephart, & Tesauro, 2001).

To succeed in such environments, agents need to make a clear distinction between their own decision-making models, the models others may be using to make decisions, and the extent to which agents deviate from these models when they actually make their decisions. This paper contributes a language, called Networks of Influence Diagrams (NID), that makes explicit the different mental models agents may use to make their decisions. NIDs provide for a clear and compact representation with which to reason about agents' beliefs and their decision-making processes. It allows multiple possible mental models of deliberation for agents, with uncertainty over which models agents are using. It is recursive, so that the mental model for an agent may itself contain models of the mental models of other agents, with associated uncertainty. In addition, NIDs allow agents' beliefs to form cyclic structures, of the form, "I believe that you believe that I believe,...", and this cycle is explicitly represented in the language. NIDs can also describe agents' conflicting beliefs about each other. For example, one can describe a scenario in which two agents disagree about the beliefs or behavior of a third agent.

NIDs are a graphical language whose building blocks are Multi Agent Influence Diagrams (MAID) (Koller & Milch, 2001). Each mental model in a NID is represented by a MAID, and the models are connected in a (possibly cyclic) graph. Any NID can be converted to an equivalent MAID that will represent the subjective beliefs of each agent in the game.

We provide an equilibrium definition for NIDs that combines the normative aspects of decision-making (what agents *should* do) with the descriptive aspects of decision-making (what agents are *expected* to do). The equilibrium makes an explicit distinction between two types of strategies: Optimal strategies represent agents' best course of action given their beliefs over others. Descriptive strategies represent how agents may deviate from their optimal strategy. In the classical approach to game theory, the normative aspect (what agents should do) and the descriptive aspect (what analysts or other agents expect them to do), have coincided. Identification of these two aspects makes sense when an agent can do no better than optimize its decisions relative to its own model of the world. However, in open environments, it is important to consider the possibility that an agent is deviating from its rational strategy with respect to its model.

NIDs share a relationship with the Bayesian game formalism, commonly used to model uncertainty over agents' payoffs in economics (Harsanyi, 1967). In this formalism, there is a type for each possible payoff function an agent may be using. Although NIDs are representationally equivalent to Bayesian games, we argue that they are a more compact, succinct and natural representation. Any Bayesian game can be converted to a NID in linear time. Any NID can be converted to a Bayesian game, but the size of the Bayesian game may be exponential in the size of the NID.

This paper is a revised and expanded version of previous work (Gal & Pfeffer, 2003a, 2003b, 2004), and is organized as follows: Section 2 presents the syntax of the NID language, and shows how they build on MAIDs in order to express the structure that holds between agents' beliefs. Section 3 presents the semantics of NIDs in terms of MAIDs, and provides an equilibrium definition for NIDs. Section 4 provides a series of examples illustrating the representational benefits of NIDs. It shows how agents can construct belief hierarchies of each other's decision-making in order to represent agents' conflicting or incorrect belief structures, cyclic belief structures and opponent modeling. It also shows how certain forms of bounded rationality can be modeled by making a distinction between agents' models of deliberation and the way they behave in reality. Section 5 demonstrates how NIDs can model "I believe that you believe" type reasoning in practice. It describes a NID that was able to outperform the top programs that were submitted to a competition for automatic rock-paper-scissors players, whose strategy was not known in advance. Section 6 compares NIDs to several existing formalisms for describing uncertainty over decision-making processes. It provides a linear time algorithm for converting Bayesian games to NIDs. Finally, Section 7 concludes and presents future work.

## 2. NID Syntax

The building blocks of NIDs are Bayesian networks (Pearl, 1988), and Multi Agent Influence Diagrams (Koller & Milch, 2001). A Bayesian network is a directed acyclic graph in which each node represents a random variable. An edge between two nodes $X_1$ and $X_2$ implies that $X_1$ has a direct influence on the value of $X_2$. Let $\mathbf{Pa}(X_i)$ represent the set of parent nodes for $X_i$ in the network. Each node $X_i$ contains a conditional probability distribution (CPD) over its domain for any value of its parents, denoted $P(X_i \mid \mathbf{Pa}(X_i))$. The topology of the network describes the conditional independence relationships that hold in the domain — every node in the network is conditionally independent of its non-descendants given its parent nodes. A Bayesian network defines a complete joint probability distribution over its random variables that can be decomposed as the product of the conditional probabilities of each node given its parent nodes. Formally,

$$P(X_1, \ldots, X_n) = \prod_{i=1}^{n} P(X_i \mid \mathbf{Pa}(X_i))$$

We illustrate Bayesian networks through the following example.

**Example 2.1.** Consider two baseball team managers Alice and Bob whose teams are playing the late innings of a game. Alice, whose team is hitting, can attempt to advance a runner by instructing him to "steal" a base while the next pitch is being delivered. A successful steal will result in a benefit to the hitting team and a loss to the pitching team, or it may result in the runner being "thrown out", incurring a large cost to the hitting team and a benefit to the pitching team. Bob, whose team is pitching, can instruct his team to throw a "pitch out", thereby increasing the probability that a stealing runner will be thrown out. However, throwing a pitch out incurs a cost to the pitching team. The decisions whether to steal and pitch out are taken simultaneously by both team managers. Suppose that the game is not tied, that is either Alice's or Bob's team is leading in score, and that the identity of the leading team is known to Alice and Bob when they make their decision.

Suppose that Alice and Bob are using pre-specified strategies to make their decisions described as follows: when Alice is leading, she instructs a steal with probability 0.75, and Bob calls a pitch out with probability 0.90; when Alice is not leading, she instructs a steal with probability 0.65, and Bob calls a pitch out with probability 0.50. There are six random variables in this domain: **Steal** and **PitchOut** represent the decisions for Alice and Bob; **ThrownOut** represents whether the runner was thrown out; **Leader** represents the identity of the leading team; **Alice** and **Bob** represent the utility functions for Alice and Bob. Figure 1 shows a Bayesian network for this scenario.

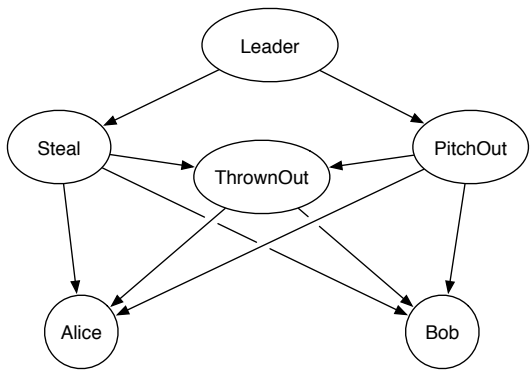

Figure 1: Bayesian network for Baseball Scenario (Example 2.1)

The CPD associated with each node in the network represents a probability distribution over its domain for any value of its parents. The CPDs for nodes **Leader**, **Steal**, **PitchOut**, and **ThrownOut** in this Bayesian network are shown in Table 1. For example, the CPD for **ThrownOut**, shown in Table 1d, represents the conditional probability distribution $P(\textsf{ThrownOut} \mid \textsf{Steal}, \textsf{PitchOut})$. According to the CPD, when Alice instructs a runner to steal a base there is an 80% chance to get thrown out when Bob calls a pitch out and a 60% chance to get thrown out when Bob remains idle. The nodes **Alice** and **Bob** have deterministic CPDs, assigning a utility for each agent for any joint value of the parent nodes **Leader**, **Steal**, **PitchOut** and **ThrownOut**. The utility for Alice is shown in Table 2. The utility for Bob is symmetric and assigns the negative value assigned by Alice's utility for the same value of the parent nodes. For example, when Alice is leading, and she instructs a runner to steal a base, Bob instructs a pitch out, and the runner is thrown out, then Alice incurs a utility of −60, while Bob incurs a utility of 60.[1]

---

1. Note that when Alice does not instruct to steal base, the runner cannot be thrown out, and the utility for both agents is not defined for this case.

**(a) node Leader**

| Leader | | |
|---|---|---|
| *alice* | *bob* | *none* |
| 0.4 | 0.3 | 0.3 |

**(b) node Steal**

| | Steal | |
|---|---|---|
| Leader | *true* | *false* |
| *alice* | 0.75 | 0.25 |
| *bob* | 0.65 | 0.35 |

**(c) node PitchOut**

| | PitchOut | |
|---|---|---|
| Leader | *true* | *false* |
| *alice* | 0.90 | 0.10 |
| *bob* | 0.50 | 0.50 |

**(d) node ThrownOut**

| | | ThrownOut | |
|---|---|---|---|
| Steal | PitchOut | *true* | *false* |
| *true* | *true* | 0.8 | 0.2 |
| *true* | *false* | 0.6 | 0.4 |
| *false* | *true* | 0 | 1 |
| *false* | *false* | 0 | 1 |

Table 1: Conditional Probability Tables (CPDs) for Bayesian network for Baseball Scenario (Example 2.1)

| Leader | Steal | PitchOut | ThrownOut | Alice |
|---|---|---|---|---|
| *alice* | *true* | *true* | *true* | −60 |
| *alice* | *true* | *true* | *false* | 110 |
| *alice* | *true* | *false* | *true* | −80 |
| *alice* | *true* | *false* | *false* | 110 |
| *alice* | *false* | *true* | *true* | — |
| *alice* | *false* | *true* | *false* | 10 |
| *alice* | *false* | *false* | *true* | — |
| *alice* | *false* | *false* | *false* | 0 |
| *bob* | *true* | *true* | *true* | −90 |
| *bob* | *true* | *true* | *false* | 110 |
| *bob* | *true* | *false* | *true* | −100 |
| *bob* | *true* | *false* | *false* | 110 |
| *bob* | *false* | *true* | *true* | — |
| *bob* | *false* | *true* | *false* | 20 |
| *bob* | *false* | *false* | *true* | — |
| *bob* | *false* | *false* | *false* | 0 |

Table 2: Alice's utility (Example 2.1) (Bob's utility is symmetric, and assigns negative value to Alice's value).

## 2.1 Multi-agent Influence Diagrams

While Bayesian networks can be used to specify that agents play specific strategies, they do not capture the fact that agents are free to choose their own strategies, and they cannot be analyzed to compute the optimal strategies for agents. Multi-agent Influence Diagrams (MAID), address these issues by extending Bayesian networks to strategic situations, where agents must choose the values of their decisions to maximize their own utilities, contingent on the fact that other agents are choosing the values of their decisions to maximize their own utilities. A MAID consists of a directed graph with three types of nodes: Chance nodes, drawn as ovals, represent choices of nature, as in Bayesian networks. Decision nodes, drawn as rectangles, represent choices made by agents. Utility nodes, drawn as diamonds, represent agents' utility functions. Each decision and utility node in a MAID is associated with a particular agent. There are two kinds of edges in a MAID: Edges leading to chance and utility nodes represent probabilistic dependence, in the same manner as edges in a Bayesian network. Edges leading into decision nodes represent information that is available to the agents at the time the decision is made. The domain of a decision node represents the choices that are available to the agent making the decision. The parents of decision nodes are called *informational parents*. There is a total ordering over each agent's decisions, such that earlier decisions and their informational parents are always informational parents of later decisions. This assumption is known as *perfect recall* or *no forgetting*. The CPD of a chance node specifies a probability distribution over its domain for each value of the parent nodes, as in Bayesian networks. The CPD of a utility node represents a deterministic function that assigns a probability of 1 to the utility incurred by the agent for any value of the parent nodes.

In a MAID, a strategy for decision node $D_i$ maps any value of the informational parents, denoted as $\mathbf{pa}_i$, to a choice for $D_i$. Let $\mathbf{C}_i$ be the domain of $D_i$. The choice for the decision can be any value in $\mathbf{C}_i$. A *pure* strategy for $D_i$ maps each value of the informational parents to an action $c_i \in \mathbf{C}_i$. A *mixed* strategy for $D_i$ maps each value of the informational parents to a distribution over $\mathbf{C}_i$. Agent $\alpha$ is free to choose any mixed strategy for $D_i$ when it makes that decision. A strategy profile for a set of decisions in a MAID consists of strategies specifying a complete plan of action for all decisions in the set.

The MAID for Example 2.1 is shown in Figure 2. The decision nodes Steal and PitchOut represent Alice's and Bob's decisions, and the nodes Alice and Bob represent their utilities. The CPDs for the chance node Leader and ThrownOut are as described in Tables 1a and 1d.

A MAID definition does not specify strategies for its decisions. These need to be computed or assigned by some process. Once a strategy exists for a decision, the relevant decision node in the MAID can be converted to a chance node that follows the strategy. This chance node will have the same domain and parent nodes as the domain and informational parents for the decision node in the MAID. The CPD for the chance node will equal the strategy for the decision. We then say that the chance node in the Bayesian network *implements* the strategy in the MAID. A Bayesian network represents a complete strategy profile for the MAID if each strategy for a decision in the MAID is implemented by a relevant chance node in the Bayesian network. We then say that the Bayesian network implements that strategy profile. Let $\sigma$ represent the strategy profile that implements all decisions in the MAID. The distribution defined by this Bayesian network is denoted by $P^\sigma$.

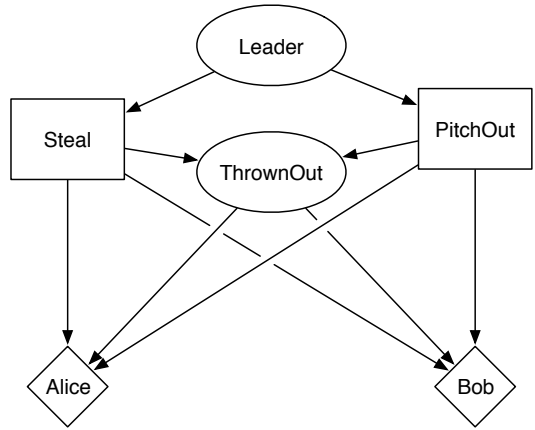

Figure 2: MAID for Baseball Scenario (Example 2.1)

An agent's utility function is specified as the aggregate of its individual utilities; it is the sum of all of the utilities incurred by the agent in all of the utility nodes that are associated with the agent.

**Definition 2.2.** Let $\mathbf{E}$ be a set of observed nodes in the MAID representing evidence that is available to $\alpha$ and let $\sigma$ be a strategy profile for all decisions. Let $\mathbf{U}(\alpha)$ be the set of all utility nodes belonging to $\alpha$. The expected utility for $\alpha$ given $\sigma$ and $\mathbf{E}$ is defined as

$$U^\sigma(\alpha \mid \mathbf{E}) = \sum_{U \in \mathbf{U}(\alpha)} E^\sigma[U \mid \mathbf{E}] = \sum_{U \in \mathbf{U}(\alpha)} \sum_{u \in \text{Dom}(U)} P^\sigma(u \mid \mathbf{E}) \cdot u$$

Solving a MAID requires computing an optimal strategy profile for all of the decisions, as specified by the Nash equilibrium for the MAID, defined as follows.

**Definition 2.3.** A strategy profile $\sigma$ for all decisions in the MAID is a Nash equilibrium if each strategy component $\sigma_i$ for decision $D_i$ belonging to agent $\alpha$ in the MAID is one that maximizes the utility achieved by the agent, given that the strategy for other decisions is $\sigma_{-i}$.

$$\sigma_i \in \operatorname*{argmax}_{\tau_i \in \Delta \mathbf{S}_i} U^{\langle \tau_i, \sigma_{-i} \rangle}(\alpha) \tag{1}$$

These equilibrium strategies specify what each agent should do at each decision given the available information at the decision. When the MAID contains several sequential decisions, the no-forgetting assumption implies that these decisions can be taken sequentially by the agent, and that all previous decisions are available as observations when the agent reasons about its future decisions.

Any MAID has at least one Nash equilibrium. Exact and approximate algorithms have been proposed for solving MAIDs efficiently, in a way that utilizes the structure of the network (Koller & Milch, 2001; Vickrey & Koller, 2002; Koller, Meggido, & von Stengel, 1996; Blum, Shelton, & Koller, 2006). Exact algorithms for solving MAIDs decompose the MAID graph into subsets of interrelated sub-games, and then proceed to find a set of equilibria in these sub-games that together constitute a global equilibrium for the entire game. In the case that there are multiple Nash equilibria, these algorithms will select one of them, arbitrarily. The MAID in Figure 2 has a single Nash equilibrium, which we can obtain by solving the MAID: When Alice is leading, she instructs her runner to steal a base with probability 0.2, and remain idle with probability 0.8, while Bob calls a pitch out with probability 0.3, and remains idle with probability 0.7. When Bob is leading, Alice instructs a steal with probability 0.8, and Bob calls a pitch out with probability 0.5.

The Bayesian network that implements the Nash equilibrium strategy profile for the MAID can be queried to predict the likelihood of interesting events. For example, we can query the network in Figure 2 and find that the probability that the stealer will get thrown out, given that agents' strategies follow the Nash equilibrium strategy profile, is 0.57.

Any MAID can be converted to an extensive form game — a decision tree in which each vertex is associated with a particular agent or with nature. Splits in the tree represent an assignment of values to chance and decision nodes in the MAID; leaves of the tree represent the end of the decision-making process, and are labeled with the utilities incurred by the agents given the decisions and chance node values that are instantiated along the edges in the path leading to the leaf. Agents' imperfect information regarding the actions of others are represented by the set of vertices they cannot tell apart when they make a particular decision. This set is referred to as an *information set*. Let $D$ be a decision in the MAID belonging to agent $\alpha$. There is a one-to-one correspondence between values of the informational parents of $D$ in the MAID and the information sets for $\alpha$ at the vertices representing its move for decision $D$.

## 2.2 Networks of Influence Diagrams

To motivate NIDs, consider the following extension to Example 2.1.

**Example 2.4.** Suppose there are experts who will influence whether or not a team should steal or pitch out. There is social pressure on the managers to follow the advice of the experts, because if the managers' decision turns out to be wrong they can assign blame to the experts. The experts suggest that Alice should call a steal, and Bob should call a pitch out. This advice is common knowledge between the managers. Bob may be uncertain as to whether Alice will in fact follow the experts and steal, or whether she will ignore them and play a best-response with respect to her beliefs about Bob. To quantify, Bob believes that with probability 0.7, Alice will follow the experts, while with probability 0.3, Alice will play best-response. Alice's beliefs about Bob are symmetric to Bob's beliefs about Alice: With probability 0.7 Alice believes Bob will follow the experts and call a pitch out, and with probability 0.3 Alice believes that Bob will play the best-response strategy with respect to his beliefs about Alice. The probability distribution for other variables in this example remains as shown in Table 1.

NIDs build on top of MAIDs to explicitly represent this structure. A *Network of Influence Diagrams (NID)* is a directed, possibly cyclic graph, in which each node is a MAID. To avoid confusion with the internal nodes of each MAID, we will call the nodes of a NID *blocks*. Let $D$ be a decision belonging to agent $\alpha$ in block $K$, and let $\beta$ be any agent. (In particular, $\beta$ may be agent $\alpha$ itself.) We introduce a new type of node, denoted Mod$[\beta, D]$ with values that range over each block $L$ in the NID. When Mod$[\beta, D]$ takes value $L$, we say that agent $\beta$ in block $K$ is *modeling* agent $\alpha$ as using block $L$ to make decision $D$. This means that $\beta$ believes that $\alpha$ may be using the strategy computed in block $L$ to make decision $D$. For the duration of this paper, we will refer to a node Mod$[\beta, D]$ as a "Mod node" when agent $\beta$ and decision $D$ are clear from context.

A Mod node is a chance node just like any other; it may influence, or be influenced by other nodes of $K$. It is required to be a parent of the decision $D$ but it is not an informational parent of the decision. This is because an agent's strategy for $D$ does not specify what to do for each value of the Mod node. Every decision $D$ will have a Mod$[\beta, D]$ node for each agent that makes a decision in block $K$, including agent $\alpha$ itself that owns the decision. If the CPD of Mod$[\beta, D]$ assigns positive probability to some block $L$, then we require that $D$ exists in block $L$ either as a decision node or as a chance node. If $D$ is a chance node in $L$, this means that $\beta$ believes that agent $\alpha$ is playing like an automaton in $L$, using a fixed, possibly mixed strategy for $D$; if $D$ is a decision node in $L$, this means that $\beta$ believes $\alpha$ is analyzing block $L$ to determine the course of action for $D$. For presentation purposes, we also add an edge $K \rightarrow L$ to the NID, labeled $\{\beta, D\}$.

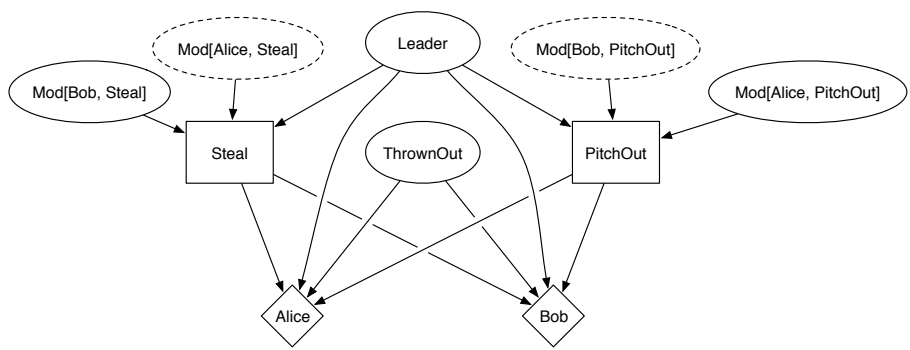

(a) *Top-level* Block

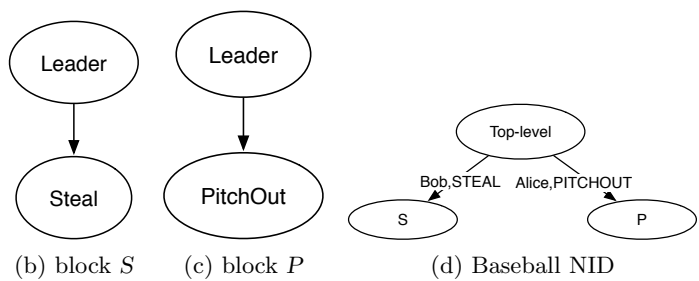

(b) block $S$      (c) block $P$      (d) Baseball NID

Figure 3: Baseball Scenario (Example 2.1)

We can represent Example 2.4 in the NID described in Figure 3. There are three blocks in this NID. The *Top-level* block, shown in Figure 3a, corresponds to an interaction between Alice and Bob in which they are free to choose whether to steal base or call a pitch out, respectively. This block is identical to the MAID of Figure 2, except that each decision node includes the Mod nodes for all of the agents. Block $S$, presented in Figure 3b, corresponds to a situation where Alice follows the expert recommendation and instructs her player to steal.

| Mod[Bob, Steal] | | Mod[Alice, PitchOut] | | Mod[Bob, PitchOut] | Mod[Alice, Steal] |
|---|---|---|---|---|---|
| *Top-level* | *S* | *Top-level* | *P* | *Top-level* | *Top-level* |
| 0.3 | 0.7 | 0.3 | 0.7 | 1 | 1 |

(a) node
Mod[Bob, Steal]

(b) node
Mod[Alice, PitchOut]

(c) node
Mod[Bob, PitchOut]

(d) node
Mod[Alice, Steal]

Table 3: CPDs for *Top-level* block of NID for Baseball Scenario (Example 2.1)

In this block, the Steal decision is replaced with a chance node, which assigns probability 1 to *true* for any value of the informational parent Leader. Similarly, block $P$, presented in Figure 3c, corresponds to a situation where Bob instructs his team to pitch out. In this block, the PitchOut decision is replaced with a chance node, which assigns probability 1 to *true* for any value of the informational parent Leader.

The root of the NID is the *Top-level* block, which in this example corresponds to reality. The Mod nodes in the *Top-level* block capture agents' beliefs over their decision-making processes. The node Mod[Bob, Steal] represents Bob's belief about which block Alice is using to make her decision Steal. Its CPD assigns probability 0.3 to the *Top-level* block, and 0.7 to block *S*. Similarly, the node Mod[Alice, PitchOut] represents Alice's beliefs about which block Bob is using to make the decision PitchOut. Its CPD assigns probability 0.3 to the *Top-level* block, and 0.7 to block *P*. These are shown in Table 3.

An important aspect of NIDs is that they allow agents to express uncertainty about the block they themselves are using to make their own decisions. The node Mod[Alice, Steal] in the *Top-level* block represents Alice's beliefs about which block Alice herself is using to make her decision Steal. In our example, the CPD of this node assigns probability 1 to the *Top-level*. Similarly, the node Mod[Bob, PitchOut] represents Bob's beliefs about which block he is using to make his decision PitchOut, and assigns probability 1 to the *Top-level* block. Thus, in this example, both Bob and Alice are uncertain about which block the other agent is using to make a decision, but not about which block they themselves are using.

However, we could also envision a situation in which an agent is unsure about its own decision-making. We say that if Mod[$\beta, D$] at block $K$ equals some block $L \neq K$, and $\beta$ owns decision $D$, then agent $\beta$ is *modeling itself* as using block $L$ to make decision $D$. In Section 3.2 we will show how this allows to capture interesting forms of bounded rational behavior. We *do* impose the requirement that there exists no cycle in which each edge includes a label $\{\alpha, D\}$. In other words, there is no cycle in which the same agent is modeling itself at each edge. Such a cycle is called a *self-loop*. This is because the MAID representation for a NID with a self-loop will include a cycle between the nodes representing the agent's beliefs about itself at each block of the NID.

In future examples, we will use the following convention: If there exists a Mod[$\alpha, D$] node at block $K$ (regardless of whether $\alpha$ owns the decision) and the CPD of Mod[$\alpha, D$] assigns probability 1 to block $K$, we will omit the node Mod[$\alpha, D$] from the block description. In the *Top-level* block of Figure 3a, this means that both nodes Mod[Alice, Steal] and Mod[Bob, PitchOut], currently appearing as dashed ovals, will be omitted.

## 3. NID Semantics

In this section we provide semantics for NIDs in terms of MAIDs. We first show how a NID can be converted to a MAID. We then define a NID equilibrium in terms of a Nash equilibrium of the constructed MAID.

### 3.1 Conversion to MAIDs

The following process converts each block $K$ in the NID to a MAID fragment $O^K$, and then connects them to form a MAID representation of the NID. The key construct in this process is the use of a chance node $D_\alpha^K$ in the MAID to represent the beliefs of agent $\alpha$ regarding the action that is chosen for decision $D$ at block $K$. The value of $D_\alpha$ depends on the block used by $\alpha$ to model decision $D$, as determined by the value of the $\text{Mod}[\alpha, D]$ node.

1. For each block $K$ in the NID, we create a MAID $O^K$. Any chance or utility node $N$ in block $K$ that is a descendant of a decision node in $K$ is replicated in $O^K$, once for each agent $\alpha$, and denoted $N_\alpha^K$. If $N$ is not a descendant of a decision node in $K$, it is copied to $O^K$ and denoted $N^K$. In this case, we set $N_\alpha^K = N^K$ for any agent $\alpha$.

2. If $P$ is a parent of $N$ in $K$, then $P_\alpha^K$ will be made a parent of $N_\alpha^K$ in $O^K$. The CPD of $N_\alpha^K$ in $O^K$ will be equal to the CPD of $N$ in $K$.

3. For each decision $D$ in $K$, we create a decision node $\text{BR}[D]^K$ in $O^K$, representing the optimal action for $\alpha$ for this decision. If $N$ is a chance or decision node which is an informational parent of $D$ in $K$, and $D$ belongs to agent $\alpha$, then $N_\alpha^K$ will be made an informational parent of $\text{BR}[D]^K$ in $O^K$.

4. We create a chance node $D_\alpha^K$ in $O^K$ for each agent $\alpha$. We make $\text{Mod}[\alpha, D]^K$ a parent of $D_\alpha^K$. If decision $D$ belongs to agent $\alpha$, then we make $\text{BR}[D]^K$ a parent of $D_\alpha^K$. If decision $D$ belongs to agent $\beta \neq \alpha$, then we make $D_\beta^K$ a parent of $D_\alpha^K$.

5. We assemble all the MAID fragments $O^K$ into a single MAID $O$ as follows: We add an edge $D_\alpha^L \to D_\beta^K$ where $L \neq K$ if $L$ is assigned positive probability by $\text{Mod}[\beta, D]^K$, and $\alpha$ owns decision $D$. Note that $\beta$ may be any agent, including $\alpha$ itself.

6. We set the CPD of $D_\alpha^K$ to be a multiplexer. If $\alpha$ owns $D$ then the CPD of $D_\alpha^K$ assigns probability 1 to $\text{BR}[D]^K$ when $\text{Mod}[\alpha, D]^K$ equals $K$, and assigns probability 1 to $D_\alpha^L$ when $\text{Mod}[\alpha, D]^K$ equals $L \neq K$. If $\beta \neq \alpha$ owns $D$ then the CPD of $D_\alpha^K$ assigns probability 1 to $D_\beta^K$ when $\text{Mod}[\alpha, D]^K$ equals $K$, and assigns probability 1 to $D_\beta^L$ when $\text{Mod}[\alpha, D]^K$ equals $L \neq K$.

To explain, Step 1 of this process creates a MAID fragment $O^K$ for each NID block. All nodes that are ancestors of decision nodes — representing events that occur prior to the decisions — are copied to $O^K$. However, events that occur *after* decisions are taken may depend on the actions for those decisions. Every agent in the NID may have its own beliefs about these actions and the events that follow them, regardless of whether that agent owns the decision. Therefore, all of the descendant nodes of decisions are duplicated for each agent in $O^K$. Step 2 ensures that if any two nodes are connected in the original block $K$, then the nodes representing agents' beliefs in $O^K$ are also connected. Step 3 creates a decision node in $O^K$ for each decision node in block $K$ belonging to agent $\alpha$. The informational parents for the decision in $O^K$ are those nodes that represent the beliefs of $\alpha$ about its informational parents in $K$. Step 4 creates a separate chance node in $O^K$ for each agent $\alpha$ that represents its belief about each of the decisions in $K$. If $\alpha$ owns the decision, this node depends on the decision node belonging to $\alpha$. Otherwise, this node depends on the beliefs of $\alpha$ regarding the action of agent $\beta$ that owns the decision. In the case that $\alpha$ models $\beta$ as using a different block to make the decisions, Step 5 connects between the MAID fragments of each block. Step 6 determines the CPDs for the nodes representing agents' beliefs about each other's decisions. The CPD ensures that the block that is used to model a decision is determined by the value of the Mod node. The MAID that is obtained as a result of this process is a complete description of agents' beliefs over each other's decisions.

We demonstrate this process by converting the NID of Example 2.4 to its MAID representation, shown in Figure 4. First, MAID fragments for the three blocks *Top-level*, $P$, and $S$ are created. The node Leader appearing in blocks *Top-level*, $P$, and $S$ is not a descendant of any decision. Following Step 1, it is created once in each of the MAID fragments, giving the nodes $\mathsf{Leader}^{TL}$, $\mathsf{Leader}^P$ and $\mathsf{Leader}^S$. Similarly, the node Steal in block $S$ and the node PitchOut in block $P$ are created once in each MAID fragment, giving the nodes $\mathsf{Steal}^S$ and $\mathsf{PitchOut}^P$. Also in Step 1, the nodes $\mathrm{Mod[Alice, Steal]}^{TL}$, $\mathrm{Mod[Bob, Steal]}^{TL}$, $\mathrm{Mod[Alice, PitchOut]}^{TL}$ and $\mathrm{Mod[Bob, PitchOut]}^{TL}$ are added to the MAID fragment for the *Top-level* block.

Step 3 adds the decision nodes $\mathrm{BR}^{TL}[\mathsf{Steal}]$ and $\mathrm{BR}^{TL}[\mathsf{PitchOut}]$ to the MAID fragment for the *Top-level* block. Step 4 adds the chance nodes $\mathsf{PitchOut}^{TL}_{\mathrm{Bob}}$, $\mathsf{PitchOut}^{TL}_{\mathrm{Alice}}$, $\mathsf{Steal}^{TL}_{\mathrm{Alice}}$ and $\mathsf{Steal}^{TL}_{\mathrm{Bob}}$ to the MAID fragment for the *Top-level* block. These nodes represent agents' beliefs in this block about their own decisions or the decisions of other agents. For example, $\mathsf{PitchOut}^{TL}_{\mathrm{Bob}}$ represents Bob's beliefs about its decision whether to pitch out, while $\mathsf{PitchOut}^{TL}_{\mathrm{Alice}}$ represents Alice's beliefs about Bob's beliefs about this decision. Also following Step 4, edges $\mathrm{BR}^{TL}[\mathsf{PitchOut}] \rightarrow \mathsf{PitchOut}^{TL}_{\mathrm{Bob}}$ and $\mathsf{Steal}^{TL}_{\mathrm{Alice}} \rightarrow \mathsf{Steal}^{TL}_{\mathrm{Bob}}$ are added to the MAID fragment for the *Top-level* block. These represent Bob's beliefs over its own decision at the block. An edge $\mathsf{Steal}^{TL}_{\mathrm{Alice}} \rightarrow \mathsf{Steal}^{TL}_{\mathrm{Bob}}$ is added to the MAID fragment to represent Bob's beliefs over Alice's decision at the *Top-level* block. There are also nodes representing Alice's beliefs about her and Bob's decisions in this block.

In Step 5, edges $\mathsf{Steal}^S \rightarrow \mathsf{Steal}^{TL}_{\mathrm{Bob}}$ and $\mathsf{PitchOut}^P \rightarrow \mathsf{Pitchout}^{TL}_{\mathrm{Alice}}$ are added to the MAID fragment for the *Top-level* block. This is to allow Bob to reason about Alice's decision in block $S$, and for Alice to reason about Bob's decision in block $P$. This action unifies the MAID fragments into a single MAID. The parents of $\mathsf{Steal}^{TL}_{\mathrm{Bob}}$ are $\mathrm{Mod[Bob, Steal]}^{TL}$, $\mathsf{Steal}^S$ and $\mathsf{Steal}^{TL}_{\mathrm{Alice}}$. Its CPD is a multiplexer node that determines Bob's prediction about Alice's action: If $\mathrm{Mod[Bob, Steal]}^{TL}$ equals $S$, then Bob believes Alice to be using block $S$, in which her action is to follow the experts and play strategy $\mathsf{Steal}^S$. If $\mathrm{Mod[Bob, Steal]}^{TL}$ equals the *Top-level* block, then Bob believes Alice to be using the *Top-level* block, in which Alice's action is to respond to her beliefs about Bob. The situation is similar for Alice's decision $\mathsf{Steal}^{TL}_{\mathrm{Alice}}$ and the node $\mathrm{Mod[Alice, Steal]}^{TL}$ with the following exception: When $\mathrm{Mod[Alice, Steal]}^{TL}$ equals the *Top-level* block, then Alice's action follows her decision node $\mathrm{BR}^{TL}[\mathsf{Steal}]$.

In the Appendix, we prove the following theorem.

**Theorem 3.1.** *Converting a NID into a MAID will not introduce a cycle in the resulting MAID.*

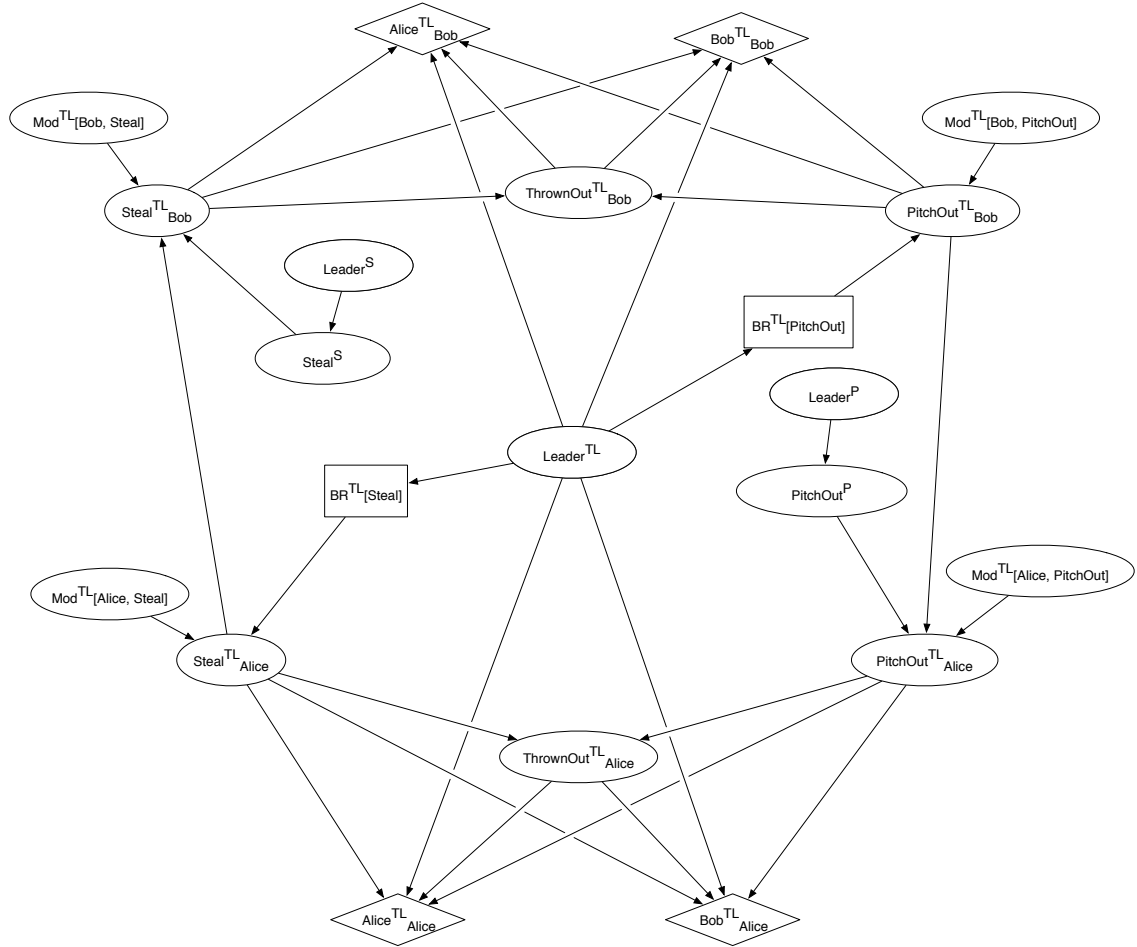

Figure 4: MAID representation for the NID of Example 2.4

As this conversion process implies, NIDs and MAIDs are equivalent in their expressive power. However, NIDs provide several advantages over MAIDs. A NID block structure makes explicit agents' different beliefs about decisions, chance variables and utilities in the world. It is a mental model of the way agents reason about decisions in the block. MAIDs do not distinguish between the real world and agents' mental models of the world or of each other, whereas NIDs have a separate block for each mental model. Further, in the MAID, nodes simply represent agents' chance, decision or utilities, and are not inherently interpreted in terms of beliefs. A $D_\alpha^K$ node in a MAID representation for a NID does not inherently represent agent $\alpha$'s beliefs about how decision $D$ is made in mental model $K$, and the $Mod^K$ for agent $\alpha$ does not inherently represent which mental model is used to make a decision. Indeed, there are no mental models defined in a MAID. In addition, there is no relationship in a MAID between descendants of decisions $N_\alpha^K$ and $N_\beta^K$, so there is no sense in which they represent the possibly different beliefs of agents $\alpha$ and $\beta$ about $N$.

Together with the NID construction process described above, a NID is a blueprint for constructing a MAID that describes agents' mental models. Without the NID, this process becomes inherently difficult. Furthermore, the constructed MAID may be large and unwieldy compared to a NID block. Even for the simple NID of Example 2.4, the MAID of Figure 4 is complicated and hard to understand.

## 3.2 Equilibrium Conditions

In Section 2.1, we defined pure and mixed strategies for decisions in MAIDs. In NIDs, we associate the strategies for decisions with the blocks in which they appear. A pure strategy for a decision $D$ in a NID block $K$ is a mapping from the informational parents of $D$ to an action in the domain of $D$. Similarly, a mixed strategy for $D$ is a mapping from the informational parents of $D$ to a distribution over the domain of $D$. A strategy profile for a NID is a set of strategies for all decisions at all blocks in the NID.

Traditionally, an equilibrium for a game is defined in terms of best response strategies. A Nash equilibrium is a strategy profile in which each agent is doing the best it possibly can, given the strategies of the other agents. Classical game theory predicts that all agents will play a best response. NIDs, on the other hand, allow us to describe situations in which an agent deviates from its best response by playing according to some other decision-making process. We would therefore like an equilibrium to specify not only what the agents should do, but also to predict what they actually do, which may be different.

A NID equilibrium includes two types of strategies. The first, called a *best response* strategy, describes what the agents should do, given their beliefs about the decision-making processes of other agents. The second, called an *actually played* strategy, describes what agents will actually do according to the model described by the NID. These two strategies are mutually dependent. The best response strategy for a decision in a block takes into account the agent's beliefs about the actually played strategies of all the other decisions. The actually played strategy for a decision in a block is a mixture of the best response for the decision in the block, and the actually played strategies for the decision in other blocks.

**Definition 3.2.** Let $\mathbf{N}$ be a NID and let $\mathbf{M}$ be the MAID representation for $\mathbf{N}$. Let $\sigma$ be an equilibrium for $\mathbf{M}$. Let $D$ be a node belonging to agent $\alpha$ in block $K$ of $\mathbf{N}$. Let the parents of $D$ be $\mathbf{Pa}$. By the construction of the MAID representation detailed in Section 3.1, the parents of $\mathrm{BR}[D]^K$ in $\mathbf{M}$ are $\mathbf{Pa}_\alpha^K$ and the domains of $\mathbf{Pa}$ and $\mathbf{Pa}_\alpha^K$ are the same. Let $\sigma_{\mathrm{BR}[D]^K}(\mathbf{pa})$ denote the mixed strategy assigned by $\sigma$ for $\mathrm{BR}[D]^K$ when $\mathbf{Pa}_\alpha^K$ equals $\mathbf{pa}$. The *best response* strategy for $D$ in $K$, denoted $\theta_D^K(\mathbf{pa})$, defines a function from values of $\mathbf{Pa}$ to distributions over $D$ that satisfy

$$\theta_D^K(\mathbf{pa}) \equiv \sigma_{\mathrm{BR}[D]^K}(\mathbf{pa})$$

In other words, the best response strategy is the same as the MAID equilibrium when the corresponding parents take on the same values.

**Definition 3.3.** Let $P^\sigma$ denote the distribution that is defined by the Bayesian network that implements $\sigma$. The *actually played* strategy for decision $D$ in $K$ that is owned by agent $\alpha$, denoted $\phi_D^K(\mathbf{pa})$, specifies a function from values of $\mathbf{Pa}$ to distributions over $D$ that satisfy

$$\phi_D^K(\mathbf{pa}) \equiv P^\sigma(D_\alpha^K \mid \mathbf{pa})$$

Note here, that $D_\alpha^K$ is conditioned on the informational parents of decision $D$ rather than its own parents. This node represents the beliefs of $\alpha$ about decision $K$. Therefore, the actually played strategy for $D$ in $K$ represents $\alpha$'s belief about $D$ in $K$, given the informational parents of $D$.

**Definition 3.4.** Let $\sigma$ be a MAID equilibrium. The NID equilibrium corresponding to $\sigma$ consists of two strategy profiles $\theta$ and $\phi$, such that for every decision $D$ in every block $K$, $\theta_D^K$ is the best response strategy for $D$ in $K$, and $\phi_D^K$ is the actually played strategy for $D$ in $K$.

For example, consider the constructed MAID for our baseball example in Figure 4. The best response strategies in the NID equilibrium specify strategies for the nodes Steal and PitchOut in the *Top-level* block that belong to Alice and Bob respectively. For an equilibrium $\sigma$ of the MAID, the best response strategy for Steal in the *Top-level* block is the strategy specified by $\sigma$ for $\mathrm{BR}^{TL}[\mathsf{Steal}]$. Similarly, the best response strategy for Pitchout in the *Top-level* block is the strategy specified by $\sigma$ for $\mathrm{BR}^{TL}[\mathsf{Pitchout}]$. The actually played strategy for Steal in the *Top-level* is equal to the conditional probability distribution over $\mathsf{Steal}_{\mathrm{Alice}}^{TL}$ given the informational parent $\mathsf{Leader}^{TL}$. Similarly, the actually played strategy for Pitchout is equal to the conditional probability distribution over $\mathsf{Pitchout}_{\mathrm{Bob}}^{TL}$ given the informational parent $\mathsf{Leader}^{TL}$. Solving this MAID yields the following unique equilibrium: In the NID *Top-level* block, the CPD for nodes Mod[Alice, Steal] and Mod[Bob, Pitchout] assigns probability 1 to the *Top-level* block, so the actually played and best response strategies for Bob and Alice are equal and specified as follows: If Alice is leading, then Alice steals base with probability 0.56 and Bob pitches out with probability 0.47. If Bob is leading, then Alice never steals base and Bob never pitches out. It turns out that because the experts may instruct Bob to call a pitch out, Alice is considerably less likely to steal base, as compared to her equilibrium strategy for the MAID of Example 2.1, where none of the managers considered the possibility that the other was being advised by experts. The case is similar for Bob.

A natural consequence of this definition is that the problem of computing NID equilibria reduces to that of computing MAID equilibria. Solving the NID requires to convert it to its MAID representation and solving the MAID using exact or approximate solution algorithms. The size of the MAID is bounded by the size of a block times the number of blocks times the number of agents. The structure of the NID can then be exploited by a MAID solution algorithm (Koller & Milch, 2001; Vickrey & Koller, 2002; Koller et al., 1996; Blum et al., 2006).

## 4. Examples

In this section, we provide a series of examples demonstrating the benefits of NIDs for describing and representing uncertainty over decision-making processes in a wide variety of domains.

### 4.1 Irrational Agents

Since the challenge to the notion of perfect rationality as the foundation of economic systems presented by Simon (1955), the theory of bounded rationality has grown in different directions. From an economic point of view, bounded rationality dictates a complete deviation from the utility maximizing paradigm, in which concepts such as "optimization" and "objective functions" are replaced with "satisficing" and "heuristics" (Gigerenzer & Selten, 2001). These concepts have recently been formalized by Rubinstein (1998). From a traditional AI perspective, an agent exhibits bounded rationality if its program is a solution to the constrained optimization problem brought about by limitations of architecture or computational resources (Russell & Wefald, 1991). NIDs serve to complement these two prevailing perspectives by allowing to control the extent to which agents are behaving irrationally with respect to their model.

Irrationality is captured in our framework by the distinction between best response and actually played strategies. Rational agents always play a best response with respect to their models. For rational agents, there is no distinction between the normative behavior prescribed for each agent in each NID block, and the descriptive prediction of how the agent actually *would* play when using that block. In this case, the best response and actually played strategies of the agents are equal. However, in open systems, or when people are involved, we may need to model agents whose behavior differs from their best response strategy. In other words, their best response strategies and actually played strategies are different. We can capture agent $\alpha$ behaving (partially) irrationally about its decision $D_\alpha$ in block $K$ by setting the CPD of $\text{Mod}[\alpha, D_\alpha]$ to assign positive probability to some block $L \neq K$.

There is a natural way to express this distinction in NIDs through the use of the Mod node. If $D_\alpha$ is a decision associated with agent $\alpha$, we can use $\text{Mod}[\alpha, D_\alpha]$ to describe which block $\alpha$ actually uses to make the decision $D_\alpha$. In block $K$, if $\text{Mod}[\alpha, D_\alpha]$ is equal to $K$ with probability 1, then it means that within $K$, $\alpha$ is making the decision according to its beliefs in block $K$, meaning that $\alpha$ will be rational; it will play a best response to the strategies of other agents, given its beliefs. If, however, $\text{Mod}[\alpha, D_\alpha]$ assigns positive probability to some block $L$ other than $K$, it means that there is some probability that $\alpha$ will not play a best response to its beliefs in $K$, but rather play a strategy according to some other block $L$. In this case, we say $\alpha$ *self-models* at block $K$. The introduction of actually played strategies into the equilibrium definition represents another advantage of NIDs over MAIDs, in that they explicitly represent strategies for agents that may deviate from their optimal strategies.

In some cases, making a decision may lead an agent to behave irrationally by viewing the future in a considerably more positive light than is objectively likely. For example, a person undergoing treatment for a disease may believe that the treatment stands a better chance of success than scientifically plausible. In the psychological literature, this effect is referred to as *motivational bias* or *positive illusion* (Bazerman, 2001). As the following example shows, NIDs can represent agents' motivational biases in a compelling way, by making Mod nodes depend on the outcome of decision nodes.

**Example 4.1.** Consider the case of a toothpaste company whose executives are faced with two sequential decisions: whether to place an advertisement in a magazine for their leading brand, and whether to increase production of the brand. Based on past analysis, the executives know that without advertising, the probability of high sales for the brand in the next quarter will be 0.5. Placing the advertisement costs money, but the probability of high sales will rise to 0.7. Increasing production of the brand will contribute to profit if sales are high, but will hurt profit if sales are low due to the high cost of storage space. Suppose now that the company executives wish to consider the possibility of motivational bias, in which placing the advertisement will inflate their beliefs about sales to be high in the next quarter to probability 0.9. This may lead the company to increase the production of the brand when it is not warranted by the market and consequently, suffer losses. The company executives wish to compute their best possible strategy for their two decisions given the fact that they attribute a motivational bias.

A NID describing this situation is shown in Figure 5c. The *Top-level* block in Figure 5a shows the situation from the point of view of reality. It includes two decisions, whether to advertise (Advertise) and whether to increase the supply of the brand (Increase). The node Sales represents the amount of sales for the brand after the decision of whether to advertise, and the node Profit represents the profit for the company, which depends on the nodes Advertise, Increase and Sales. The CPD of Sales in the *Top-level* block assigns probability 0.7 to $high$ if Advertise is $true$ and 0.5 to $high$ if Advertise is $false$, as described in Table 4a. The utility values for node Profit are shown in Table 4.1. For example, when the company advertises the toothpaste, increases its supply, and sales are high, it receives a reward of 70; when the company advertises the toothpaste, does not increase its supply, and sales are low, it receives a reward of $-40$. Block *Bias*, described in Figure 5b, represents the company's biased model. Here, the decision to advertise is replaced by an automaton chance node that assigns probability 1 to Advertise = $true$. The CPD of Sales in block *Bias* assigns probability 0.9 to $high$ if Advertise is $true$ and 0.5 to $high$ if Advertise is $false$, as described in Table 4b. In the *Top-level* block, we have the following:

1. The node Mod[Company, Advertise] assigns probability 1 to the *Top-level* block.

2. The decision node Advertise is a parent of the node Mod[Company, Increase].

3. The node Mod[Company, Increase] assigns probability 1 to block *Bias* when Advertise is $true$, and assigns probability 0 to block *Bias* when Advertise is $false$.

Intuitively, Step 1 captures the company's beliefs that it is not biased before it makes the decision to advertise. Step 2 allows the company's uncertainty about whether it is biased to depend on the decision to advertise. Note that this example shows when it is necessary for a decision node to depend on an agent's beliefs about a past decision. Step 3 captures the company's beliefs that it may use block *Bias* to make its decision whether to increase supply, in which it is over confident about high sales.

Solving this NID results in the following unique equilibrium: In block *Bias*, the company's actually played and best response strategy is to increase supply, because this is its optimal action when it advertises and sales are high. In block *Top-level*, we have the following: If the company chooses not to advertise, it will behave rationally, and its best response and actually played strategy will be *not* to increase supply; if the company chooses to advertise, its actually played strategy will be to use block *Bias* in which it increases supply, and its best response strategy will be not to increase supply. Now, the expected utility for the company in the *Top-level* block is higher when it chooses not to advertise. Therefore, its best response strategies for both decisions are not to advertise nor to increase supply. Interestingly, if the company was never biased, it can be shown using backwards induction that its optimal action for the first decision is to advertise. Thus, by reasoning about its own possible irrational behavior for the second decision, the company revised its strategy for the first decision.

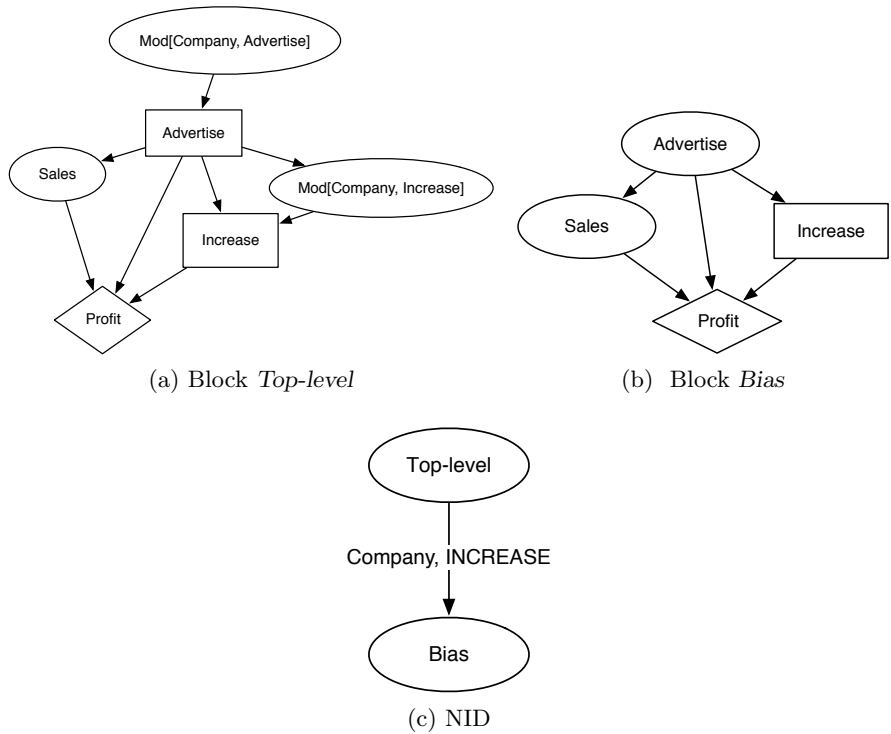

(a) Block *Top-level*

(b) Block *Bias*

(c) NID

Figure 5: Motivational Bias Scenario (Example 4.1)

|            | Sales |      |
| :---:      | :---: | :---: |
| Advertise  | *low* | *high* |
| *true*     | 0.3   | 0.7  |
| *false*    | 0.5   | 0.5  |

(a) node **Sales** (*Top-level* Block)

|            | Sales |      |
| :---:      | :---: | :---: |
| Advertise  | *low* | *high* |
| *true*     | 0.1   | 0.9  |
| *false*    | 0.5   | 0.5  |

(b) node **Sales** (*Bias* Block)

Table 4: CPDs for *Top-level* block of Motivational Bias NID (Example 4.1)

**Example 4.2.** Consider the following extension to Example 2.4. Suppose that there are now two successive pitches, and on each pitch the managers have an option to steal or pitch out. If Bob pitches out on the first pitch, his utility for pitching out on the second pitch (regardless of Alice's action) decreases by 20 units because he has forfeited two pitches. Bob believes that with probability 0.3, he will succumb to social pressure during the second pitch and call a pitch out. Bob would like to reason about this possibility when making the decision for the first pitch.

| Advertise | Increase | Sales | Profit |
|-----------|----------|-------|--------|
| *true* | *true* | *high* | 70 |
| *true* | *true* | *low* | −70 |
| *true* | *false* | *high* | 50 |
| *true* | *false* | *low* | −40 |
| *false* | *true* | *high* | 80 |
| *false* | *true* | *low* | −60 |
| *false* | *false* | *high* | 60 |
| *false* | *false* | *low* | −30 |

Table 5: Company's utility (node Profit) for *Top-level* block of Motivational Bias NID (Example 4.1)

In this example, each manager is faced with a sequential decision problem: whether to steal or pitch out in the first and second pitch. The strategy for the second pitch is relevant to the strategy for the first pitch for each agent. Now, each of the managers, if they were rational, could use backward induction to compute optimal strategies for the first pitch, by working backwards from the second pitch. However, this is only a valid procedure if the managers behave rationally on the second pitch. In the example above, Bob knows that he will be under strong pressure to pitch out on the second pitch and he wishes to take this possibility into account, while making his decision for the first pitch.

| Mod[Bob, PitchOut$_2$] | |
|------------------------|-----|
| *Top-level* | 0.7 |
| $L$ | 0.3 |

Table 6: CPD for Mod[Bob, PitchOut$_2$] node in *Top-level* block of Irrational Agent Scenario (Example 4.2)

We can model this situation in a NID as follows. The *Top-level* block of the NID is shown in Figure 6a. Here, the decision nodes Steal$_1$ and PitchOut$_1$ represent the decisions for Alice and Bob in the first pitch, and the nodes Steal$_2$ and Pitchout$_2$ represent the decisions for Alice and Bob in the second pitch. The nodes Leader, Steal$_1$, PitchOut$_1$ and ThrownOut$_1$ are all informational parents of the decision nodes Steal$_2$ and PitchOut$_2$. For expository convenience, we have not included the edges leading from node Leader to the utility nodes in the block. Block $L$, shown in Figure 6b, describes a model for the second pitch in which Bob is succumbing to social pressure and pitches out, regardless of who is leading. This is represented by having the block include a chance node PitchOut$_2$ which equals *true* with probability 1 for each value of Leader. The node Mod[Bob, PitchOut$_2$] will assign probability 0.3 to block $L$, and 0.7 probability to the *Top-level* block, as shown in Table 4.1. The node Mod[Bob, PitchOut$_2$] is not displayed in the *Top-level* block. By our convention, this implies that its CPD assigns probability 1 to the *Top-level* block, in which Bob is reasoning about the possibility of behaving irrationally with respect to the second pitch. In this way, we have captured the fact that Bob may behave irrationally with respect to the second pitch, and that he is reasoning about this possibility when making the decision for the first pitch.

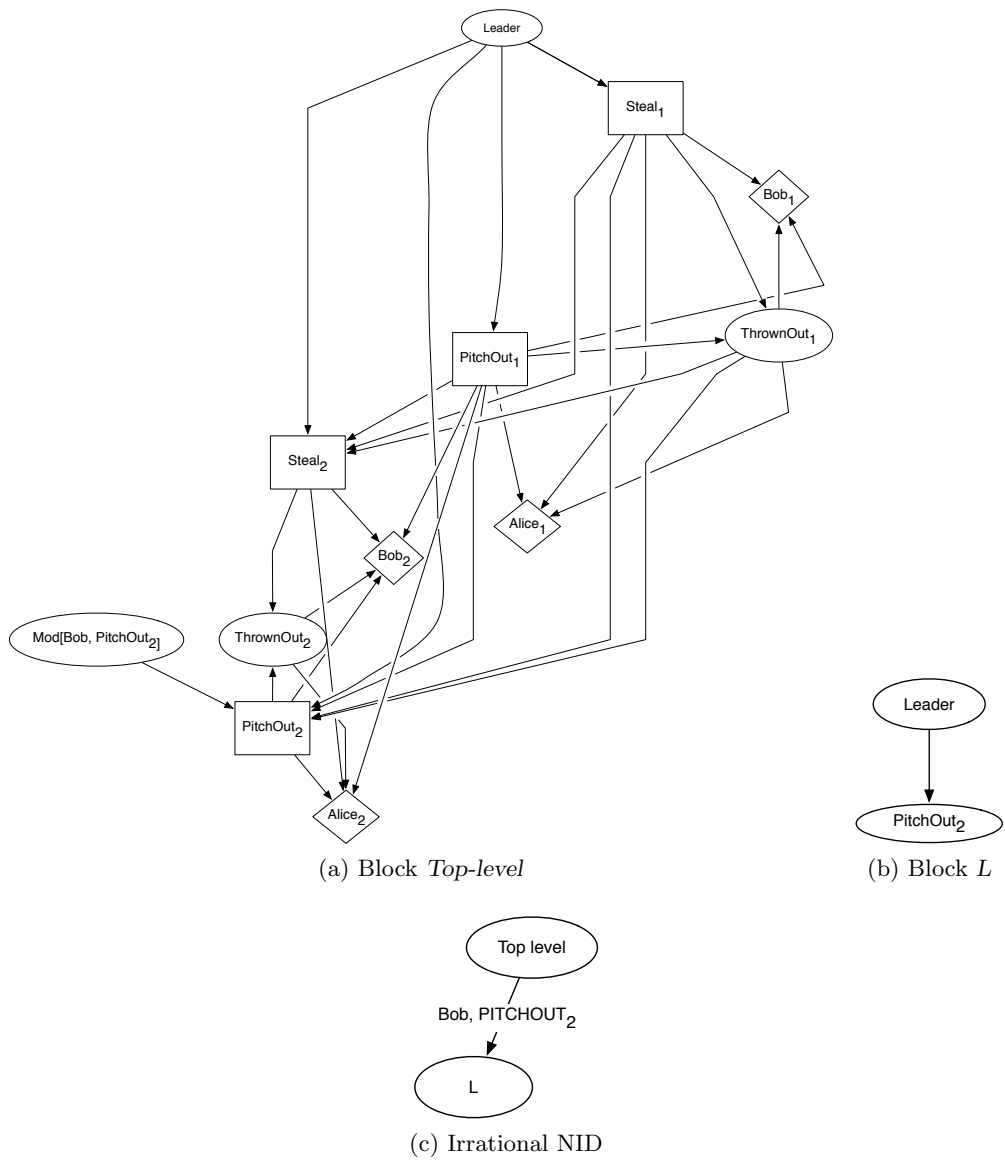

(a) Block *Top-level*

(b) Block *L*

(c) Irrational NID

Figure 6: Irrational Agent Scenario (Example 4.2)

There is a unique equilibrium for this NID. Both agents behave rationally for their first decision so their actually played and best response strategies are equal, and specified as follows: Alice steals a base with probability 0.49 if she is leading, and never steals a base if Bob is leading. Bob pitches out with probability 0.38 if Alice is leading and pitches out with probability 0.51 if Bob is leading. In the second pitch, Alice behaves rationally, and her best response and actually played strategy are as follows: steal base with probability 0.42 if Alice is leading and never steal base if Bob is leading. Bob may behave irrationally in the second pitch: His best response strategy is to pitch out with probability 0.2 if Alice is leading, and pitch out with probability 0.52 if Bob is leading; his actually played strategy is to pitch out with probability 0.58 if Alice is leading, and with probability 0.71 if Bob is leading. Note that because Bob is reasoning about his possible irrational behavior in the second pitch, he is less likely to pitch out in the first pitch as compared to the case in which Bob is completely rational (Example 2.4).

## 4.2 Conflicting Beliefs

In traditional game theory, agents' beliefs are assumed to be *consistent* with a common prior distribution, meaning that the beliefs of agents about each other's knowledge is expressed as a posterior probability distribution resulting from conditioning a common prior on each agent's information state. One consequence of this assumption is that agents' beliefs can differ only if they observe different information (Aumann & Brandenburger, 1995). This result led to theoretic work that attempted to relax the common prior assumption. Myerson (1991) showed that a game with inconsistent belief structure that is finite can be converted to a new game with *consistent* belief structures by constructing utility functions that are equivalent to the original game in a way that they both assign the same expected utility to the agents. However, this new game will include beliefs and utility functions that are fundamentally different to the original game exhibiting the inconsistent belief structure. For a summary of the economic and philosophical ramifications of relaxing the common prior assumption, see the work of Morris (1995) and Bonanno and Nehring (1999).

Once we have a language that allows us to talk about different mental models that agents have about the world, and different beliefs that they have about each other and about the structure of the game, it is natural to relax the common prior assumption within NIDs while preserving the original structure of the game.

**Example 4.3.** Consider the following extension to the baseball scenario of Example 2.1. The probability that the runner is thrown out depends not only on the decisions of both managers, but also on the *speed* of the runner. Suppose a fast runner will be thrown out with 0.4 probability when Bob calls a pitch out, and with 0.2 probability when Bob does not call a pitch out. A slow runner will be thrown out with 0.8 probability when Bob calls a pitch out, and with 0.6 probability when Bob does not call a pitch out.

Now, Bob believes the runner to be slow, but is unsure about Alice's beliefs regarding the speed of the runner. With probability 0.8, Bob believes that Alice thinks that the stealer is fast, and with probability 0.2 Bob believes that Alice thinks that the stealer is slow. Assume that the distributions for other variables in this example are as described in Table 1.

In this example, Bob is uncertain whether Alice's beliefs about the speed of the runner conflict with his own. NIDs allow to express this in a natural fashion by having two blocks that describe the same decision-making process, but differ in the CPD that they assign to the speed of the runner. Through the use of the Mod node, NIDs can specify agents' conflicting beliefs about which of the two blocks is used by Alice to make her decision, according to Bob's beliefs. The NID and blocks for this scenario are presented in Figure 7.

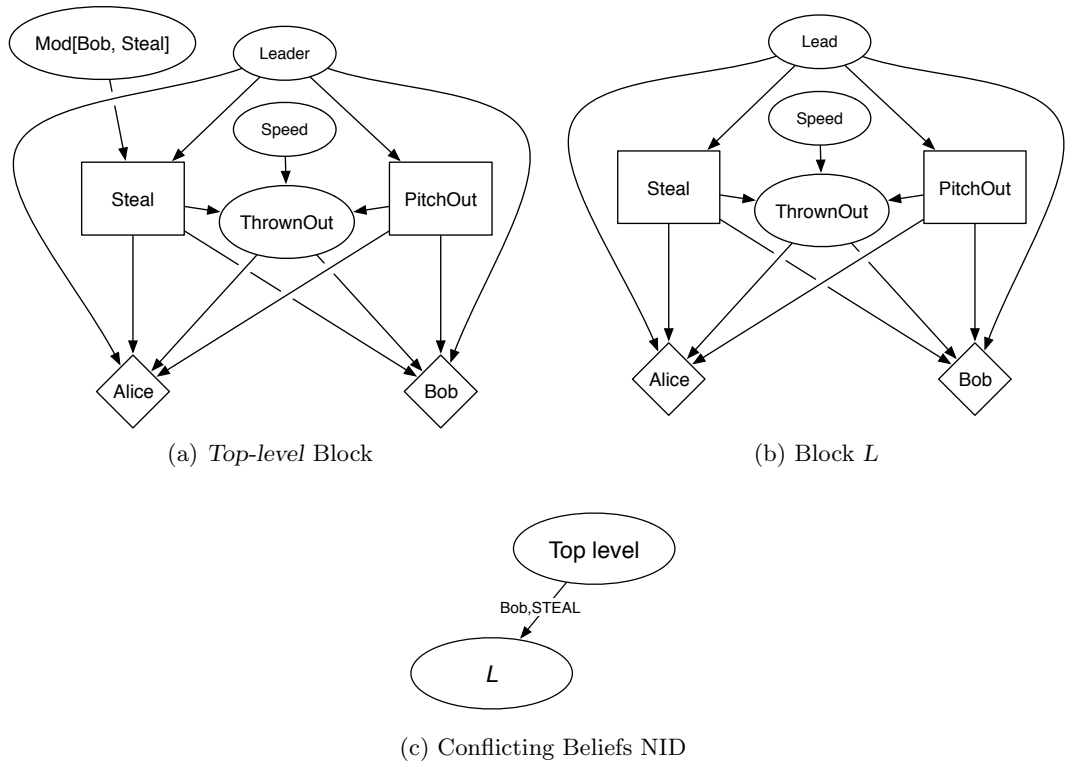

(a) *Top-level* Block

(b) Block *L*

(c) Conflicting Beliefs NID

Figure 7: Conflicting Beliefs Scenario (Example 4.3)

In the *Top-level* block, shown in Figure 7a, Bob and Alice decide whether to pitch out or to steal base, respectively. This block is identical in structure to the *Top-level* block of the previous example, but it has an additional node Speed that is a parent of node ThrownOut, representing the fact that the speed of the runner affects the probability that the runner is thrown out.

The *Top-level* corresponds to Bob's model, in which the runner is slow. The CPD of the node Speed assigns probability 1 to *slow* in this block, as shown in Table 7a. Block *L*, shown in Figure 7b, represents an identical decision-making process as in the *Top-level* block, except that the CPD of Speed is different: it assigns probability 1 to *fast*, as shown in Table 7b. The complete NID is shown in Figure 7c. Bob's uncertainty in the *Top-level* block over Alice's decision-making process is represented by the node Mod[Bob, Steal], whose CPD is shown in Table 7c. With probability 0.8, Alice is assumed to be using block *L*, in which the speed of the runner is fast. With probability 0.2, Alice is assumed to be using the *Top-level* block, in which the speed of the runner is slow. Note that in the

| Speed | |
|---|---|
| *fast* | *slow* |
| 0 | 1 |

(a) node **Speed** (block *Top-level*)

| Speed | |
|---|---|
| *fast* | *slow* |
| 1 | 0 |

(b) node **Speed** (block *L*)

| Mod[Bob, Steal] | |
|---|---|
| *Top-level* | *L* |
| 0.2 | 0.8 |

(c) node Mod[Bob, **Steal**] (block *Top-level*)

Table 7: CPDs for nodes in Conflicting Beliefs NID (Example 4.3)

*Top-level* block, the nodes Mod[Alice, **Steal**], Mod[Alice, **PitchOut**] and Mod[Bob, **PitchOut**] are not displayed. By the convention introduced earlier, all these nodes assign probability 1 to the *Top-level* block and have been omitted from the *Top-level* block of Figure 7a. Interestingly, this implies that Alice knows the runner to be slow, even though Bob believes that Alice believes the runner is fast. When solving this NID, we get a unique equilibrium. Both agents are rational, so their best response and actually played strategies are equal, and specified as follows: In block *L*, the runner is fast, so Alice always steals base, and Bob always calls a pitch out. In the *Top-level* block, Bob believes that Alice uses block *L* with high probability, in which she seals a base. In the *Top-level* block the speed of the runner is slow and will likely be thrown out. Therefore, Bob does not pitch out in order to maximize its utility given its beliefs about Alice. In turn, Alice does not steal base at the *Top-level* block because the speed of the runner is slow at this block.

## 4.3 Collusion and Alliances

In a situation where an agent is modeling multiple agents, it may be important to know whether those agents are working together in some fashion. In such situations, the models of how the other agents make their decisions may be correlated, due to possible collusion.

**Example 4.4.** A voting game involves 3 agents Alice, Bob, and Carol, who are voting one of them to be chairperson of a committee. Alice is the incumbent, and will be chairperson if the vote ends in a draw. Each agent would like itself to be chairperson, and receives utility 2 in that case. Alice also receives a utility of 1 if she votes for the winner but loses the election, because she wants to look good. Bob and Carol, meanwhile, dislike Alice and receive utility -1 if Alice wins.

It is in the best interests of agents Bob and Carol to coordinate, and both vote for the same person. If Bob and Carol do indeed coordinate, it is in Alice's best interest to vote for the person they vote for. However, if Bob and Carol mis-coordinate, Alice should vote for herself to remain the chairperson. In taking an opponent modeling approach, Alice would like to have a model of how Bob and Carol are likely to vote. Alice believes that with probability 0.2, Bob and Carol do not collude; with probability 0.3, Bob and Carol collude to vote for Bob; with probability 0.4, Bob and Carol collude to vote for Carol. Also, Alice believes that when they collude, both agents might renege and vote for themselves with probability 0.1.

This example can easily be captured in a NID. The *Top-level* block is shown in Figure 8. There is a node **Collude**, which will have three possible values: none indicating no collusion;

Bob and Carol indicating collusion to vote for Bob or Carol respectively. The decision nodes A, B, C represent the decisions for Alice, Bob and Carol, respectively. The CPD for Collude is presented in Table 8a. The nodes Mod[Alice, B] and Mod[Alice, C], whose CPD is shown in Table 8b and 8c respectively, depend on Collude. If Collude is none, Mod[Alice, B] will assign probability 1 to the *Top-level* block. If Collude is Bob, Mod[Alice, B] will equal a block *B* describing an automaton in which Bob and Carol both vote for Bob. If Collude is Carol, Mod[Alice, B] will equal a block *C*, in which Bob and Carol vote for Carol with probability 0.9, and block *B* with probability 0.1. This accounts for the possibility that when Bob and Carol have agreed to vote for Carol, Bob might renege. The CPD for Mod[Alice, B] is similar, and is described in Table 8b. The CPD for Mod[Alice, C] is symmetric, and is described in Table 8c.

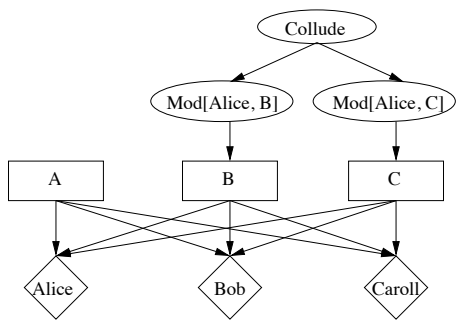

Figure 8: *Top-level* block of Collusion Scenario (Example 4.4)

|  | Collude |  |  |
|---|---|---|---|
|  | none | Bob | Carol |
|  | 0.2 | 0.3 | 0.5 |

(a) node Collude

| Mod[Alice, *B*] | Collude | | |
|---|---|---|---|
|  | none | Bob | Carol |
| *Top-level* | 1 | 0 | 0 |
| *B* | 0 | 1 | 0 |
| *C* | 0 | 0.1 | 0.9 |

(b) node Mod[Alice, *B*]

| Mod[Alice, *C*] | Collude | | |
|---|---|---|---|
|  | none | Bob | Carol |
| *Top-level* | 1 | 0 | 0 |
| *B* | 0 | 0.9 | 0.1 |
| *C* | 0 | 0 | 1 |

(c) node Mod[Alice, *C*]

Table 8: CPDs for *Top-level* block of Collusion Scenario (Example 4.4)

In the unique NID equilibrium for this example, all agents are rational so their actually played and best response strategies are equal. In the equilibrium, Alice always votes for Carol because she believes that Bob and Carol are likely to collude and vote for Carol.

In turn, Carol votes for herslef or for Bob with probability 0.5, and Bob always votes for himself. By reneging, Bob gives himself a chance to win the vote, in the case that Carol votes for him.

Moving beyond this example, one of the most important issues in multi-player games is alliances. When players form an alliance, they will act for the benefit of the alliance rather than purely for their own self-interest. Thus an agent's beliefs about the alliance structure affects its models of how other agents make their decisions. When an agent has to make a decision in such a situation, it is important to be able to model its uncertainty about the alliance structure.

### 4.4 Cyclic Belief Structures

Cyclic belief structures are important in game theory, where they are used to model agents who are symmetrically modeling each other. They are used to describe an infinite regress of "I think that you think that I think..." reasoning. Furthermore, cyclic belief structures can be expressed in economic formalisms, like Bayesian games, so it is vital to allow them in NIDs in order for NIDs to encompass Bayesian games. Cyclic belief structures can naturally be captured in NIDs by including a cycle in the NID graph.

**Example 4.5.** Recall Example 4.3, in which Alice and Bob had conflicting beliefs about the speed of the runner. Suppose that Bob believes that the runner is slow, and that with probability 0.8, Alice believes that the runner is fast, and is modeling Bob as reasoning about Alice's beliefs, and so on...

We model this scenario using the cyclic NID described in Figure 9c. In the *Top-level* block, shown in Figure 9b, Bob believes the runner to be slow and is modeling Alice as using block $L$ to make her decision. In block $L$, Alice believes the runner to be fast, and is modeling Bob as using the *Top-level* block to make his decision. Bob's beliefs about Alice in the *Top-level* block are represented by the CPD of node Mod[Bob, Steal], shown in Table 9c, which assigns probability 1 to block $L$.

In block $L$, the CPD of Speed, shown in Table 9b assigns probability 1 to $fast$. Alice's beliefs about Bob in block $L$ are represented by the CPD of node Mod[Alice, PitchOut], shown in Table 9d, which assigns probability 1 to block $L$. In the *Top-level* block, the CPD of Speed assigns probability 1 to *slow*, shown in Table 4.4a. The NID equilibrium for this scenario is as follows. In both blocks $L$ and *Top-level*, Alice does not steal base, and Bob does not pitch out, regardless of who is leading.

## 5. Application: Opponent Modeling

In some cases, agents use rules, heuristics, patterns or tendencies when making decisions. One of the main approaches to game playing with imperfect information is opponent modeling, in which agents try to learn the patterns exhibited by other players and react to their model of others. NIDs provide a solid, coherent foundation for opponent modeling.

**Example 5.1.** In the game of RoShamBo (commonly referred to as Rock-Paper-Scissors), players simultaneously choose one of *rock*, *paper*, or *scissors*. If they choose the same item, the result is a tie; otherwise rock crushes scissors, paper covers rock, or scissors cut paper, as shown in Table 10.

| Speed | |
|-------|------|
| *fast* | *slow* |
| 0 | 1 |

(a) node Speed
(block *Top-level*)

| Speed | |
|-------|------|
| *fast* | *slow* |
| 1 | 0 |

(b) node Speed
(block *L*)

| Mod[Bob, Steal] | |
|-----------------|---|
| *Top-level* | *L* |
| 1 | 0 |

(c) node
Mod[Bob, Steal]
(block *Top-level*)

| Mod[Alice, PitchOut] | |
|----------------------|---|
| *Top-level* | *L* |
| 1 | 0 |

(d) node
Mod[Alice, PitchOut]
(block *L*)

Table 9: CPDs for nodes in Cyclic NID (Example 4.5)

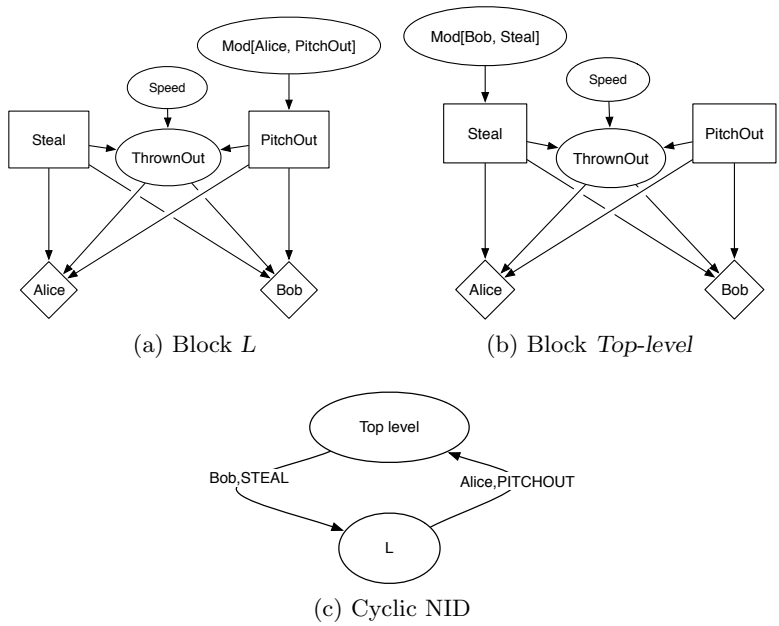

(a) Block *L*

(b) Block *Top-level*

(c) Cyclic NID

Figure 9: Cyclic Baseball Scenario (Example 4.5)

| | *rock* | *paper* | *scissors* |
|----------|---------|---------|------------|
| *rock* | (0, 0) | (−1, 1) | (1, −1) |
| *paper* | (1, −1) | (0, 0) | (−1, 1) |
| *scissors* | (−1, 1) | (1, −1) | (0, 0) |

Table 10: Payoff Matrix for Rock-paper-scissors

The game has a single Nash equilibrium in which both players play a mixed strategy over $\{rock, paper, scissors\}$ with probability $\{\frac{1}{3}, \frac{1}{3}, \frac{1}{3}\}$. If both players do not deviate from their equilibrium strategy, they are guaranteed an expected payoff of zero. In fact, it is easy to verify that a player who always plays his equilibrium strategy is guaranteed to get an expected zero payoff regardless of the strategy of his opponent. In other words, sticking to the equilibrium strategy guarantees not to lose a match in expectation, but it also guarantees not to win it!

However, a player can try and win the game if the opponents are playing suboptimally. Any suboptimal strategy can be beaten, by predicting the next move of the opponent and then employing a counter-strategy. The key to predicting the next move is to model the strategy of the opponent, by identifying regularities in its past moves.

Now consider a situation in which two players play repeatedly against each other. If a player is able to pick up the tendencies of a suboptimal opponent, it might be able to defeat it, assuming the opponent continues to play suboptimally. In a recent competition (Billings, 2000), programs competed against each other in matches consisting of 1000 games of RoShamBo. As one might expect, Nash equilibrium players came in the middle of the pack because they broke even against every opponent. It turned out that the task of modeling the opponent's strategy can be surprisingly complex, despite the simple structure of the game itself. This is because sophisticated players will attempt to counter-model their opponents, and will hide their own strategy to avoid detection. The winning program, called Iocaine Powder (Egnor, 2000), did a beautiful job of modeling its opponents on multiple levels. Iocaine Powder considered that its opponent might play randomly, according to some heuristic, or it might try to learn a pattern used by Iocaine Powder, or it might play a strategy designed to counter Iocaine Powder learning its pattern, or several other possibilities.

## 5.1 A NID for Modeling Belief Hierarchies

Inspired by "Iocaine Powder", we constructed a NID for a player that is playing a match of RoShamBo and is trying to model his opponent. Suppose that Bob wishes to model Alice's play using a NID. The block *Top-level* of the NID, shown in Figure 10a, is simply a MAID depicting a RoShamBo round between Bob and Alice. Both players have access to a predictor P, an algorithm that is able to predict the next move in a sequence as a probability distribution over the possible moves. The only information available to the predictor is the history of past moves for Alice and Bob.

Alice may be ignoring P, and playing the Nash Equilibrium strategy. Bob has several alternative models of Alice's decision. According to block *Automaton*, shown in Figure 10c, Alice always follows the signal P. In block *B1*, shown in Figure 10b, Bob is modeling Alice as using block *Automaton* to make her decision. This is achieved by setting the CPD of Mod[Bob, Alice] in block *B1* to assign probability 1 to *Automaton*. We can analyze the NID rooted at block *B1* to determine Bob's best response to Alice. For example, if Bob thinks, based on the history, that P is most likely to tell Alice to play *rock*, then Bob would play *paper*. Let us denote this strategy as $BR(\mathsf{P})$.

However, Alice can also model Bob by assigning probability 1 to Mod[Alice, Bob] in block *A1*. In this way, Alice is reasoning about Bob modeling Alice as following the predictor P.

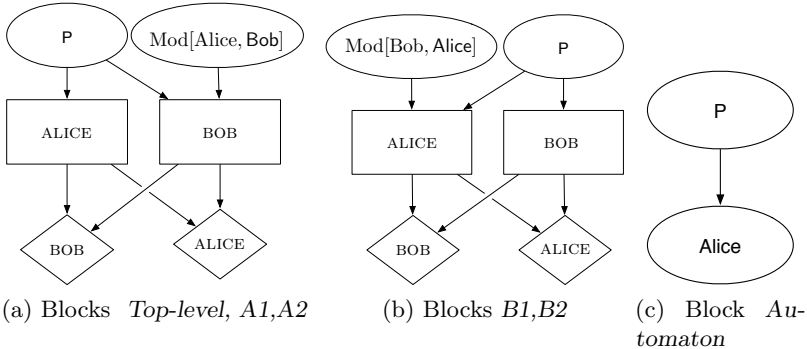

(a) Blocks *Top-level, A1,A2*

(b) Blocks *B1,B2*

(c) Block *Automaton*

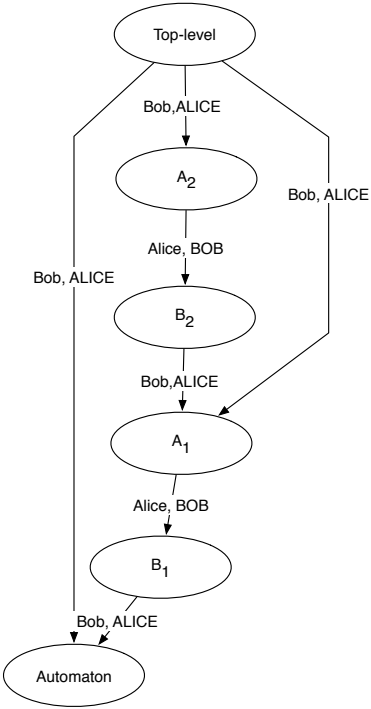

(d) RoShamBo NID

Figure 10: RoShamBo Scenario (Example 5.1)

When we analyze the NID originating in block A1, shown in Figure 10a, we will determine Alice's best-response to Bob's model of her as well as Bob's best-response to his model of Alice. Since Alice believes that Bob plays $BR(\mathsf{P})$ as a result of Bob's belief that Alice plays according to $\mathsf{P}$, she will therefore play a best response to $BR(\mathsf{P})$, thereby double-guessing Bob. Alice's strategy in block *A1* is denoted as $BR(BR(\mathsf{P}))$. Following our example, in block *A1* Alice does not play *rock* at all, but *scissors*, in order to beat Bob's play of *paper*. Similarly, in block *B2*, Bob models Alice as using block *A1* to make her decisions, and in block *A2*, Alice models Bob as using block *B2* to make his decision. Therefore, solving the NID originating in block *B2* results in a $BR(BR(BR(\mathsf{P})))$ strategy for Bob. This would prompt Bob to play *rock* in *B2* in our example, in order to beat *scissors*. Lastly, solving the NID originating in block *A2* results in a $BR(BR(BR(BR(\mathsf{P}))))$ strategy for Alice. This would prompt Alice to play *paper* in block *A2*, in order to beat *rock*. Thus, we have shown that for every instance of the predictor $\mathsf{P}$, Alice might play one of the three possible strategies. Any pure strategy can only choose between *rock*, *paper*, or *scissors* for any given $\mathsf{P}$, so this reasoning process terminates.

The entire NID is shown in Figure 10d. In block *Top-level*, Bob models Alice as using one of several possible child blocks: block *Automaton*, in which Alice follows her predictor; block *A1*, in which Alice is second-guessing her predictor; or block *A2*, in which Alice is triple-guessing her predictor. Bob's uncertainty over Alice's decision-making processes is captured in the Mod[Bob, Alice] node in block *Top-level*. Analyzing the *Top-level* block of this NID will extract Bob's best response strategy given his beliefs about Alice's decision-making processes.

To use this NID in practice, it is necessary to compute the MAID equilibrium and extract Bob's best-response strategy at the *Top-level* block. To this end, we need to estimate the values of the NID parameters, represented by the unknown CPDs at each of its blocks, and solve the NID. These parameters include Mod[Bob, Alice], representing Bob's beliefs in the *Top-level* block regarding which block Alice is using; and node $\mathsf{P}$, representing the distributions governing the signals for Alice and Bob, respectively.[2] To this end, we use an on-line version of the the EM algorithm that was tailored for NIDs. We begin with random parameter assignments to the unknown CPDs. We then revise the estimate over the parameters of the NID given the observations at each round. Then Bob plays the best-response strategy of the MAID representation for the NID given the current parameter setting. Interleaving learning and using the NID to make a decision helps Bob to adapt to Alice's possibly changing strategy.

## 5.2 Empirical Evaluation

We evaluated the NID agent against the ten top contestants from the first automatic RoShamBo competition. All of these agents used an opponent modeling approach, that is, they learned some signal of their opponent's play based on the history of prior rounds. Contestants can be roughly classified according to three dimensions: the type of signal used (probabilistic vs. deterministic); the type of reasoning used (pattern vs. meta-reasoners); and, their degree of exploration versus exploitation of their model. Probabilistic agents estimated a distribution over the strategies of their opponents while deterministic agents predicted their opponents' next move with certainty. Pattern reasoners directly modeled their opponents as playing according to some rule or distribution, and did not reason about the possibility that their opponents were modeling themselves. In contrast, meta-reasoners attempted to double- or triple-guess their opponents' play. Exploitative agents played a best response to their model of their opponents, while explorative agents deviated, under certain conditions, from their best response strategy to try and learn different behavioral patterns of their opponents. Iocaine Powder used the strategy of reverting to the Nash equilibrium when it was losing. Because this made it impossible to evaluate whether our NID model could learn Iocaine powder's reasoning process, we turned off this strategy. Also, we limited all contestants' strategies to depend on the last 100 rounds of play, in order to allow a fair comparison with the NID agent that only used four rounds of play. We did not limit them to four rounds because they were not originally designed to use such a short history. Our purpose was to show that explicitly reasoning and learning about mental models can make a difference, and not to optimize learning the model of the signal.

---

2. Technically, the CPDs for the nodes representing prior history are also missing. However, they are observed at each decision-making point in the interaction and their CPDs do not affect players' utilities.

Figure 11 shows the performance of the RoShamBo NID when playing 10 matches of 3,000 rounds with each contestant. The overall standings were determined by ordering the total scores for each contestant in all rounds played (+1 for winning a round against a contestant by the NID player; −1 for losing a round; 0 for ties). Therefore, it was important for each player to maximize its win against the weaker opponents, and minimize its loss to stronger opponents. The $x$-axis includes the contestant number while the $y$-axis describes the difference between the average score of the RoShamBo NID and the contestant; error bars indicate a single standard deviation difference.

As shown by the figure, the RoShamBo NID was able to defeat all contestant in all matches, including a version of Iocaine Powder. The best performance for the NID was achieved when playing pattern reasoners that used deterministic signals (Contestants 3, 5 and 6). Each of these contestants directly predicted their opponents' play as a function of the history, without reasoning about their opponents' model of themselves. Consequently, it was difficult for them to detect change in the strategies of adaptive opponents, such as the RoShamBo NID. In addition, the use of deterministic signals made it harder for these contestants to capture probabilistic players like the NID algorithm.

The RoShamBo NID also outperformed those contestants that attempted to trick their opponents, by reasoning about the possibility that the opponents are double- and triple-guessing their model (Contestants 4 and 1). This shows that the NID was able to determine the level of reasoning employed by its opponents.

## 6. Relationship with Economic Models

In this section, we describe the relationship between NIDs and several existing formalisms for representing uncertainty over decision-making processes. NIDs share a close relationship with Bayesian games (Harsanyi, 1967), a game-theoretic framework for representing uncertainty over players' payoffs. Bayesian games capture the beliefs agents have about each other as well as define an equilibrium that assigns a best response strategy for each agent given its beliefs. Bayesian games are quite powerful in their ability to describe belief hierarchies and cyclic belief structures.

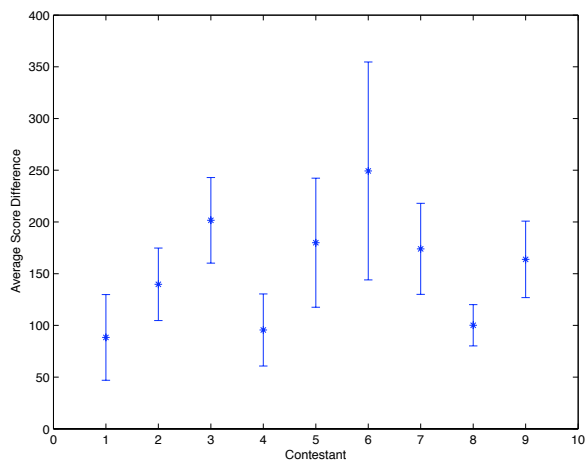

| Opponent type | Number |
|---|---|
| Iocaine Powder | 1 |
| Probabilistic, Pattern, Exploitative | 2, 9 |
| Deterministic, Pattern, Exploitative | 3, 6, 5 |
| Probabilistic, Meta, Exploitative | 1, 4 |
| Probabilistic, Pattern, Exploitative | 7, 8 |

Figure 11: Difference in average outcomes between NID player and opponents

In a Bayesian game, each agent has a discrete type embodying its private information. Let $\mathbf{N}$ be a set of agents. For each agent $i$ a Bayesian game includes a set of possible types $\mathbf{T}_i$, a set of possible actions $\mathbf{C}_i$, a conditional distribution $p_i$ and a utility function $u_i$. Let $\mathbf{T} = \times_{i \in \mathbf{N}} \mathbf{T}_i$ and let $\mathbf{C} = \times_{i \in \mathbf{N}} \mathbf{C}_i$. For each agent $i$, let $\mathbf{T}_{-i} = \times_{j \neq i} \mathbf{T}_j$ denote the set of all possible types other than those of agent $i$. The probability distribution $p_i$ is a function from $t_i$ to $\Delta \mathbf{T}_{-i}$, that is, $p_i(.|t_i)$ specifies for each type $t_i \in \mathbf{T}_i$ a joint distribution over the types of the other agents. The utility function $u_i$ is a function from $\mathbf{C} \times \mathbf{T}$ to the real numbers. It is a standard assumption that the game, including agents' strategies, utilities and type distributions, is common knowledge to all agents.

The solution concept most commonly associated with Bayesian games is a *Bayesian Nash equilibrium*. This equilibrium maps each type to a mixed strategy over its actions that is the agent's best response to the strategies of the other agents, given its beliefs about their types. Notice that in a Bayesian game, an agent's action can depend on its own types but not on the types of the other agents, because they are unknown to that agent when it analyzes the game. It is assumed that each agent knows its own type, and that this type subsumes all of the agent's private information before the game begins. Because the types of other agents are unknown, each agent maximizes its expected utility given its distribution over other types.

Let $\mathbf{N}_{-i}$ denote all of the agents in the Bayesian game apart from agent $i$. Let $\sigma_i(.|t_i)$ denote a random strategy for agent $i$ given that its type is $t_i$. A Bayesian Nash equilibrium is any mixed strategy profile $\sigma$ such that for any agent $i$ and type $t_i \in \mathbf{T}_i$ we have

$$\sigma_i(.|t_i) \in \operatorname{argmax}_{\tau \in \Delta C_i} \sum_{\mathbf{t}_{-i} \in \mathbf{T}_{-i}} p_i(\mathbf{t}_{-i}|t_i) \cdot \\ \sum_{\mathbf{c} \in \mathbf{C}} \left( \prod_{j \in N_{-i}} \sigma_j(c_j|t_j) \right) \tau(c_i) u_i(\mathbf{t}, \mathbf{c}) \tag{2}$$

Bayesian games have been used extensively for modeling interaction in which agents have private information, such as auction mechanisms (Myerson, 1991) and they can be used to express uncertainty over agents' decision-making models. In general, Bayesian games are just as expressive as NIDs. As we show, any Bayesian game can be converted into a NID in time and space linear in the size of the Bayesian game. Conversely, any NID can be converted into a Bayesian game, because any NID can be converted to a MAID, which can in turn be converted to an extensive form game. The extensive form game can be converted to a normal form game which is a trivial Bayesian game with only one type per agent. However, in the worst case, the size of the extensive form game will be exponential in the number of informational parents for decision nodes in the MAID, and the size of the normal form game will be exponential in the size of the extensive form game. Of course, this is a brute force conversion; more compact conversions may be possible.

We now consider more formally the question of whether Bayesian games can be represented by NIDs. The idea is to align each type in a Bayesian game with a decision in a NID block. The resulting best response strategy for the decision in the NID equilibrium will equal the Bayes Nash equilibrium strategy for the type.

**Definition 6.1.** Let $B$ be a Bayesian game and $N$ a NID. We say that $N$ is *equivalent* to $B$ if there exists an injective mapping $f$ from types in $B$ to (block,agent) pairs in $N$, such that the following conditions hold:

1. For any Bayesian Nash equilibrium $\sigma$ of $B$, there exists a NID equilibrium of $N$, such that for every type $t_i$, if $f$ maps $t_i$ to $(K, \alpha)$, the best-response and actually-played strategies for $\alpha$ in $K$ are equal to $\sigma_i(.|t_i)$.

2. For any NID equilibrium of $N$, there exists a Bayesian Nash equilibrium $\sigma$ of $B$ such for every $(K, \alpha)$ in the image of $f$, $\sigma_i(.|t_i)$ where $t_i = f^{-1}(K, \alpha)$ is equal to the best-response and actually-played strategies for $\alpha$ in K.

The following theorem is proved in Appendix 8.

**Theorem 6.2.** *Every Bayesian game can be represented by an equivalent NID whose size is linear in the size of the Bayesian game.*

In this section, we will use the term Bayesian games to specify a representation that includes type distributions and utility functions that are presented explicitly. NIDs enjoy the same advantages over fully specified Bayesian games that graphical models typically enjoy over unstructured representations. In general, NIDs may be exponentially more compact than Bayesian games because Bayesian games require, for every type of every agent, a full joint distribution over the types of all other agents. In addition, the utility function in a Bayesian game specifies a utility for each joint combination of types and actions of every player. These distributions and utility functions are exponential in the number of players. In NIDs, because they are based on MAIDs, the type distributions can be decomposed into a product of small conditional distributions, and the utility functions can be additively decomposed into a sum of small functions that depend only on a small number of actions.

In addition, Bayesian games are representationally obscure. First, types in Bayesian games are atomic entities that capture all the information available to an agent in a single variable. A type is used to capture both an agent's beliefs about the way the world works (including its preferences), and its private information. For example, in poker, both the player's beliefs about the other player's tendency to bluff and her knowledge of what cards she has received are captured by a type. We believe that these two aspects are fundamentally different; one describes the actual state of the world and the other describes what is going on in a player's head. Conflating these two aspects leads to confusion. In NIDs, the two aspects are differentiated. Private information about the world is represented by informational parents, whereas mental models are represented by blocks.

Second, a type in a Bayesian game does not decompose different aspects of information into variables. Thus in poker, the hand must be represented by a single variable, whereas in NIDs it can be represented by different variables representing each of the cards. A final point is that in Bayesian games all of the uncertainty must be folded into the utility functions and the distribution over agents' types. Consider the scenario in which two agents have conflicting beliefs about a chance variable, such as in Example 4.3. In a NID, there will be a separate block for each possible mental model that differs in the CPD assignments for the chance variable. In contrast, each type in the Bayesian game would sum over the distribution over the chance variable. Looking at the Bayesian game, we would not know whether the reason for the different utility functions is because the agent has different beliefs about the chance variable, or whether it is due to different preferences of the agent.

NIDs also exhibit a relationship with more recent formalisms for *games of awareness*, in which agents may be unaware of other players' strategies or of the structure of the game (Halpern & Rego, 2006; Feinberg, 2004). A game description in this formalism shows how players' awareness about each other's strategies changes over time. A game of awareness includes a set of extensive form game descriptions, called *augmented* games, that represent an analyst's beliefs about the world, as well as separate descriptions for each game that may become true according to agents' subjective beliefs. The analyst's augmented game is considered to be the actual description of reality, while each subjective augmented game can differ from the analyst's game in agents' utility functions, their decisions, and the strategies available to agents at each of their decisions. A history for an agent in an augmented game is a sub-path in the tree leading to a node in which the agent makes a move. Awareness is modeled by a function that maps an agent-history pair in one augmented game to another augmented game which the agent considers possible given the history. Uncertainty over agents' awareness in an augmented game can be quantified by having nature choose a move in the tree leading to agents' information sets. The definition of Nash equilibrium is extended to include a set of strategies for each agent-game pairthat the agent considers to be possible, given a history and the best-response strategies used by other agents in the augmented game. This formalism can capture an analyst's model about agents' awareness as well as agents' model about their own, or other agents' awareness.

There are fundamental differences between NIDs and games of awareness. First, like Bayesian games, the equilibrium conditions for this representation do not allow for agents to deviate from their best-response strategies. Second, they require the presence of a modeler agent, that in reality, is modeling its uncertainty about levels of awareness of other agents. NIDs allow for such a modeler agent, but they do not require it. This allows to capture situations where no agent has certain knowledge of reality, such as in the Baseball NID of Example 2.4. Third, each augmented game of awareness is represented as an extensive form game, that as we have shown above, may be exponentially larger than the MAID used to represent each decision-making model in a NID. Lastly, agents' awareness over each other's strategies is just one type of reasoning that can be captured by a NID. Other types of reasoning processes were described in Section 4.

Lastly, Gmytrasiewicz and Durfee (2001) have developed a framework for representing uncertainty over decision-making using a tree structure in which the nodes consist of payoff matrices for a particular agent. Like Bayesian games, uncertainty is folded into the payoff matrices. Each agent maintains its own tree, representing its model of the decision-making processes used by other agents. Like traditional representations, this language assumes that all agents behave rationally. In addition, it assumes that each agent believes others to use a fixed strategy, that is folded into the environment.

## 7. Conclusion

We have presented a highly expressive language for describing agents' beliefs and decision-making processes in games. Our language is graphical. A model in our language is a network of interrelated models, where each mental model itself is a graphical model of a game. An agent in one mental model may believe that another agent (or possibly itself) uses a different mental model to make decisions; it may have uncertainty about which mental model is used. We presented semantics for our language in terms of multi-agent influence diagrams. We analyzed the relationship between our language and Bayesian games. They are equally expressive, but NIDs may be exponentially more compact.

We showed how our language can be used to describe agents who play irrationally, in the sense that their actual play does not correspond to the best possible response given their beliefs about the world and about other agents. This is captured by a novel equilibrium concept that captures the interaction between what agents should do and what they actually do. We also showed how to express situations in which agents have conflicting beliefs, including situations in which the agents do not have a common prior distribution over the state of the world. Finally, we showed how to capture cyclic reasoning patterns, in which agents engage in infinite chains of "I think that you think that I think..." reasoning.

A vital question is the use of our language to learn about agents' behavior and reasoning processes. As we have shown, our language can be used to learn non-stationary strategies in rock-paper-scissors. In other work, we have shown how models that were inspired by NIDs can learn people's play in negotiation games (Gal, Pfeffer, Marzo, & Grosz, 2004; Gal & Pfeffer, 2006). The focus of our continuing work will be to develop a general method for learning models in NIDs.

### Acknowledgments

Thank you very much for the useful comments provided by the anonymous reviewers and editor. Thanks to Barbara Grosz and Whitman Richards for their invaluable guidance.

Thanks to Adam Juda for reading a prior draft of this work. This work was supported by an NSF Career Award IIS-0091815 and AFOSR under contract FA9550-05-1-0321.

## 8. Appendix A

**Theorem** 3.1: *Converting a NID into a MAID will not introduce a cycle in the resulting MAID.*

*Proof.* First, let us ignore the edges added by step 5 of the construction, and focus on the MAID fragment $O^K$ constructed from a single block $K$. Since the block is acyclic, we can number the nodes of the block with integers in topological order. We now number the nodes of $O^K$ as follows. For a node $N_\alpha$ that derives from a chance or utility node $N$ in $K$, $N_\alpha$ gets the same number as $N$. A node $\mathrm{BR}[D]^K$ gets the same number as $D$. A node $D_\alpha^K$, where $\alpha$ owns $D$, gets the same number as $D$ plus 1/3. A node $D_\alpha^K$, where $\alpha$ does not own $D$, gets the same number as $D$ plus 2/3. By construction, if $P$ is a parent of $N$ in $O^K$, $P$ has a lower number than $N$.

Now let us consider the entire constructed MAID $O$. Suppose, by way of contradiction, that there is a cycle in $O$. It follows from the above argument that it must consist entirely of edges between fragments added by step 5. Since all such edges emanate from a node $D_\alpha^K$ where $\alpha$ owns $D$, and end at a node $D_\alpha^L$, all nodes in the cycle must refer to the same decision $D$, and must belong to the agent who owns $D$. Thus the cycle must be of the form $D_\alpha^{K_1}, \ldots, D_\alpha^{K_n}, D_\alpha^{K_1}$ where $\alpha$ owns $D$. Since an edge has been added from $D_\alpha^{K_i}$ to $D^{K_{i+1}}$ in $O$, $\alpha$ must be modeling itself in block $K_i$ as using block $K_{i+1}$ to make decision $D$. Therefore there is a self-loop in the NID, which is a contradiction. $\square$

**Theorem** 6.2: *Every Bayesian game can be represented as an equivalent NID whose size is linear in the size of the Bayesian game.*

*Proof.* Given a Bayesian game $\mathbf{B}$, we construct a NID $\mathbf{N}$ as follows. The set of agents in $\mathbf{N}$ is equal to the set of agents in $\mathbf{B}$. For each type $t_i$ of agent $i$ in $\mathbf{B}$ there is a corresponding block in $\mathbf{N}$ labeled $t_i$. The block $t_i$ contains a decision node $D_j$ and utility node $U_j$ for every agent $j$. $D_j$ has no informational parents. The domain of $D_j$ is the set of choices $C_j$ for agent $j$ in $\mathbf{B}$. We add a new chance node $Q_i$ in block $t_i$ whose domain is the set $\mathbf{T_{-i}}$. Each node $\mathrm{Mod}[i, D_j]$ where $j \neq i$ will have the node $Q_i$ as a parent. The parents of $U_i$ are all the decision nodes as well as the node $Q_i$. For an agent $j \neq i$, $U_j$ has only the parent $D_j$. For each agent $j$ we define a distinguished action $c_j^* \in \mathbf{C}_j$.

We set the CPD for nodes in $t_i$ as follows:

1. The CPD of $\mathrm{Mod}[i, D_i]$ assigns probability 1 to $t_i$.

2. The CPD of $Q_i$ assigns probability $p_i(\mathbf{t_{-i}} \mid t_i)$, as defined in $\mathbf{B}$, for each type profile $\mathbf{t_{-i}} \in \mathbf{T_{-i}}$.

3. The CPD of a node $\mathrm{Mod}[i, D_j]$ where $j \neq i$ assigns probability 1 to block $t_j$ when the $j$th element of the value of the parent node $Q_i$ equals $t_j$. This projects the probability distribution $Q_i$ in $\mathbf{B}$ to the node $\mathrm{Mod}[i, D_j]$ representing $i$'s beliefs about which block agent $j$ is using in the NID.

4. The CPD of $U_i$ assigns probability 1 to $u_i(\mathbf{t}, \mathbf{c})$, as defined in $\mathbf{B}$, given that $Q_i$ equals $\mathbf{t}$, and $\mathbf{D}$ equals $\mathbf{c}$.

5. The CPD of $U_j$ assigns probability 1 to utility 1 when $D_j = c_j^*$, and probability 1 to 0 otherwise.

6. The CPD of $\text{Mod}[j, D_k]$, for all $k$, when $j \neq i$, assigns probability 1 to $t_i$.

Our construction is accompanied by the injective mapping $f$ that maps a type $t_i$ to the (block,agent) pair $(t_i, i)$.

Let $\mathbf{M}$ be the constructed MAID for $\mathbf{N}$. To prove condition 1 of Definition 6.1, let $\tau$ be a Bayes Nash equilibrium of $\mathbf{B}$. For each agent, $\tau_i$ is a conditional probability distribution $\tau_i(\cdot \mid t_i)$. We define the strategy profile $\sigma$ in $\mathbf{M}$ as follows. $\sigma_{\text{BR}[D_i]^{t_i}} = \tau_i(\cdot \mid t_i)$ for decisions owned by agent $i$, and $\sigma_{\text{BR}[D_j]^{t_i}}$ assigns probability 1 to $c_j^*$ when $j \neq i$.

We claim the following:

1. $\sigma$ is a MAID equilibrium in $\mathbf{M}$, according to Definition 2.3.

2. In the resulting NID equilibrium, the best response strategy for $i$ in $t_i$ is $\tau_i(\cdot \mid t_i)$.

3. In the resulting NID equilibrium, the actually played strategy is the same as the best response strategy.

Claim 3 is true because $\text{Mod}[i, D_i]$ assigns probability 1 to $t_i$.

Note that there are no informational parents in $\mathbf{N}$. Therefore, by the definition of NID equilibrium, the best response strategy $\theta_{D_i}^{t_i} = \sigma_{\text{BR}[D_i, t_i]} = \tau_i(\cdot \mid t_i)$. Therefore, Claim 2 is true.

To prove Claim 1, note first that in block $t_i$, the utility node $U_j$, where $j \neq i$, is fully determined by $D_j$, because $D_j$ is the sole parent of $U_j$. Also, player $j$ is not self-modeling at $D_j$, because the CPD of node $\text{Mod}[j, D_j]$ assigns probability 1 to $t_i$. The same holds in $\mathbf{M}$: the decision node $\text{BR}[D_j]^{t_i}$ is the sole parent of $U_j^{t_i}$. Therefore, in any equilibrium for $\mathbf{M}$, the strategy for $\text{BR}[D_j]^{t_i}$ will assign probability 1 to the distinguished action $c_j^*$ that causes $U_j^{t_i}$ to be 1.

In block $t_i$, the CPD of $\text{Mod}[i, D_j]$ assigns probability 0 to $t_i$. This means that player $j$ is not using block $t_i$ to make its decision, according to $i$'s beliefs. Therefore, $\text{BR}[D_j]^{t_i}$ is independent of $U_i^{t_i}$, and the equilibrium strategies for $\text{BR}[D_i]^{t_i}$ are independent of the distinguished action chosen for the $\text{BR}[D_i]^{t_j}$.

By the definition of MAID equilibrium of Definition 2.3, the strategy profile $\sigma$ is an equilibrium if each $\sigma_i$ maximizes $\text{EU}^\sigma(i)$. We need to show that maximizing this is equivalent to maximizing the right hand side of Equation 2. There is a utility $U_i^{t_i}$ and decision node $\text{BR}[D_i]^{t_i}$ in every block $t_i$. Let $c_i^{t_i}$ denote a choice for agent $i$ at decision $\text{BR}[D_i]^{t_i}$ in block $t_i$. Let $t_i'$ denote a block corresponding to a different type $t_i'$ for agent $i$. Let $c_i^{t_i'}$ be a choice for the agent in decision $\text{BR}[D_i]^{t_i'}$ at block $t_i'$ and $\mathbf{c}_{-i}$ all the choices for other decisions $\text{BR}[\mathbf{D}_{-i}]^{t_i}$. By the construction of $\mathbf{M}$, $U_i^{t_i}$ is d-separated from $\text{BR}[D_i]^{t_i'}$ given $\text{BR}[D_i]^{t_i}$ and $\text{BR}[\mathbf{D}_{-i}]^{t_i}$.

As a result, we can optimize $U_i^{t_i}$ separately from all other utility nodes belonging to agent $i$, considering only $\text{BR}[D_i]^{t_i}$. We then get that the utility for $i$ in $\mathbf{M}$ given the strategy profile $\sigma$ can be written as

$$E^\sigma[U_i^{t_i}] = \sum_{c_i^{t_i}} \sigma_i^{t_i}(c_i^{t_i}) \sum_{\mathbf{c}_{-i}} \sigma_{-i}(\mathbf{c}_{-i}) \sum_{u_i^{t_i}} P^\sigma(U_i^{t_i} = u_i \mid c_i^{t_i}, \mathbf{c}_{-i}) \cdot u_i^{t_i} \tag{3}$$

We now condition on agent $i$'s beliefs about the decisions of other agents in block $t_i$. Let $\text{Mod}[i, \mathbf{D}_{-i}]^{t_i}$ denote the set of nodes $\text{Mod}[i, D_j]^{t_i}$ where $j \neq i$, and let the tuple $\mathbf{t}_{-i}$ refer to the block label profile for blocks $\mathbf{T}_{-i}$. We now obtain

$$\sum_{c_i^{t_i}} \sigma_i^{t_i}(c_i^{t_i}) \sum_{\mathbf{c}_{-i}} \sigma_{-i}(\mathbf{c}_{-i}) \sum_{\mathbf{t}_{-i}} P(\text{Mod}[i, \mathbf{D}_{-i}]^{t_i} = \mathbf{t}_{-i}) \sum_{u_i^{t_i}} (u_i^{t_i} \mid c_i^{t_i}, \mathbf{c}_{-i}, \mathbf{t}_{-i}) \cdot u_i^{t_i} \tag{4}$$

Now observe that the role of $\text{Mod}[i, \mathbf{D}_{-i}]^{t_i}$ is to determine which choices for decisions $\text{BR}[\mathbf{D}_{-i}]^{t_i}$ are relevant for the utility of player $i$. In particular, if $\text{Mod}[i, D_j]^{t_i}$ is equal to $t_j$, then it is $j$'s choice in block $t_j$ that player $i$ needs to consider when it makes decision $\text{BR}[D_i]^{t_i}$. Let $\mathbf{c}_i^{\mathbf{t}_{-i}}$ denote the relevant choices for $\text{BR}[\mathbf{D}_{-i}]^{t_i}$ when $\text{Mod}[i, \mathbf{D}_{-i}]^{t_i} = \mathbf{t}_{-i}$. Since other choice variables are irrelevant, we can marginalize them out and obtain

$$\sum_{c_i^{t_i}} \sigma_i^{t_i}(c_i^{t_i}) \sum_{\mathbf{c}_{-i}^{\mathbf{t}_{-i}}} \sigma_{-i}(\mathbf{c}_{-i}^{\mathbf{t}_{-i}}) \sum_{\mathbf{t}_{-i}} P(\text{Mod}[i, \mathbf{D}_{-i}]^{t_i} = \mathbf{t}_{-i})$$
$$\sum_{u_i^{t_i}} P^\sigma(U_i^{t_i} = u_i \mid c_i^{t_i}, \mathbf{c}_{-i}^{\mathbf{t}_{-i}}) \cdot u_i^{t_i} \tag{5}$$

Rearranging terms, we rewrite Equation 5.

$$\sum_{\mathbf{t}_{-i}} P(\text{Mod}[i, \mathbf{D}_{-i}]^{t_i} = \mathbf{t}_{-i}) \sum_{\mathbf{c}} \left( \prod_{j \neq i} \sigma_j^{t_j}(c_j^{t_j}) \right)$$
$$\sigma_i^{t_i}(c_i^{t_i}) \sum_{u_i^{t_i}} P^\sigma(U_i^{t_i} = u_i \mid c_i^{t_i}, \mathbf{c}_{-i}^{\mathbf{t}_{-i}}) \cdot u_i^{t_i} \tag{6}$$

By our construction, $P(\text{Mod}[i, \mathbf{D}_{-i}]^{t_i} = \mathbf{t}_{-i})$ is $p_i(\mathbf{t}_{-i} \mid t_i)$ as defined in $\mathbf{B}$, $\sigma_j^{t_j}(c_j^{t_j})$ is $\tau_j(c_j \mid t_j)$ as defined in $\mathbf{B}$, and $\sum_{u_i^{t_i}} P^\sigma(U_i^{t_i} = u_i \mid c_i^{t_i}, \mathbf{c}_{-i}^{\mathbf{t}_{-i}}) \cdot u_i^{t_i}$ is $u_i(\mathbf{t}, \mathbf{c})$. We therefore get

$$\sum_{\mathbf{t}_{-i}} p_i(\mathbf{t}_{-i} \mid t_i) \sum_{\mathbf{c}} \left( \prod_{j \neq i} \tau_j(c_j \mid t_j) \right) \tau_i(c_i \mid t_i) u_i(\mathbf{t}, \mathbf{c}) \tag{7}$$

Therefore $\sigma$ is a MAID equilibrium of $\mathbf{M}$ if and only if $\tau$ is a Bayesian Nash equilibrium of $\mathbf{B}$. Claim 1 is established and therefore Condition 1 of Definition 6.1 is satisfied.

Finally, to prove Condition 2, given a NID equilibrium of $\mathbf{N}$ we construct a MAID equilibrium $\sigma$ for $\mathbf{M}$ by copying the best response strategies, and then construct strategies $\tau$ for $\mathbf{B}$ in exactly the reverse manner to above. The previous reasoning applies in reverse to show that $\tau$ is a Bayes Nash equilibrium of $\mathbf{B}$ and the best response and actually played strategies for $\mathbf{N}$ are equal to $\tau$.

$\square$